\documentclass[fleqn,usenatbib]{mnras}

\usepackage{newtxtext,newtxmath}
\usepackage[T1]{fontenc}
\usepackage {threeparttable}
\DeclareRobustCommand{\VAN}[3]{#2}
\let\VANthebibliography\thebibliography
\def\thebibliography{\DeclareRobustCommand{\VAN}[3]{##3}\VANthebibliography}

\usepackage{graphicx}	% Including figure files
\usepackage{amsmath}	% Advanced maths commands
\usepackage{orcidlink} 
\usepackage[subrefformat=parens]{subcaption}
\usepackage{comment} 
\usepackage{bm}
\newcommand{\lya}{\mathrm{Ly}\alpha}
\newcommand{\dnl}{D_\mathrm{NL}(k,\mu)}
\newcommand{\dm}{D_\mathrm{M}(k,\mu)}
\newcommand{\fog}{F_\mathrm{FoG}(k,\mu)}
\newcommand{\pf}{P_F(k,\mu)}
\newcommand{\pc}{P_\times(k,\mu)}
\newcommand{\pl}{P_\mathrm{L}(k)}
\newcommand{\kmax}{k_\mathrm{max}}
\newcommand{\mpc}{h^{-1}\mathrm{Mpc}}
\newcommand{\mpcinv}{h\mathrm{Mpc}^{-1}}
\newcommand{\kvav}{k^{a_v}_v}
\newcommand{\mh}{M_h}
\newcommand{\csquared}{$\chi^2/$\text{d.o.f}}

\newcommand{\himpc}{h{\rm Mpc}^{-1}}

\newcommand{\mini}{{\it Minimum}\ }
\title[Ly-$\alpha$ forest power spectrum and its cross-correlation]{Lyman-$\alpha$ forest power spectrum and its cross-correlation with dark matter halos in different astrophysical models}

\author[K. Nakashima et al.]{
Koichiro Nakashima$^{1}$\thanks{E-mail: nakashima.koichiro.v2@s.mail.nagoya-u.ac.jp},
Atsushi J. Nishizawa$^{2,3,4}$\orcidlink{0000-0002-6109-2397}\thanks{atsushi.nishizawa@iar.nagoya-u.ac.jp},
Kentaro Nagamine$^{5,6,7,8}$\,\orcidlink{0000-0001-7457-8487},
Yuri Oku$^{5,9}$\,\orcidlink{0000-0002-5712-6865}, 
Ikkoh Shimizu$^{10}$
\\
$^{1} $Department of Physics, Nagoya University, Furocho, Chikusa, Nagoya, Aichi 464-8602, Japan,\\
$^2$ DX Center, Gifu Shotoku Gakuen University, Takakuwanishi, Yanaizu, Gifu, 501-6194, Japan\\
$^3$ Institute for Advanced Research, Nagoya University, Furocho, Chikusa, Nagoya, Aichi, 464-8602, Japan\\
$^4$ Kobayashi Maskawa Institute, Nagoya University, Furocho, Chikusa, Nagoya, Aichi, 464-8602, Japan\\
$^{5}$ Theoretical Astrophysics, Department of Earth and Space Science, Graduate School of Science, Osaka University, Toyonaka, Osaka 560-0043, Japan\\
$^{6}$ Theoretical Joint Research, Forefront Research Center, Graduate School of Science, Osaka University, Toyonaka, Osaka 560-0043, Japan\\
$^{7}$ Kavli IPMU (WPI), The University of Tokyo, 5-1-5 Kashiwanoha, Kashiwa, Chiba, 277-8583, Japan \\
$^{8}$ Department of Physics \& Astronomy, University of Nevada, Las Vegas, 4505 S. Maryland Pkwy, Las Vegas, NV 89154-4002, USA \\
$^{9}$ Center for Cosmology and Computational Astrophysics, the Institute for Advanced Study in Physics, Zhejiang University, China \\
$^{10}$ National Institute of Technology, Kagawa College, 551, Koda, Takuma-cho, Mitoyo-shi, Kagawa 769-1103, Japan  \\
}

\date{Accepted XXX. Received YYY; in original form ZZZ}

\pubyear{2015}

\begin{document}
\label{firstpage}
\pagerange{\pageref{firstpage}--\pageref{lastpage}}
\maketitle

\begin{abstract}
The $\lya$ forest, a series of HI absorption lines in the quasar spectra, is a powerful tool for probing the large-scale structure of the intergalactic medium. Its three-dimensional (3D) correlation and cross-correlations with quasars allow precise measurements of the baryon acoustic oscillation feature and redshift space distortions at redshifts $z>2$. 
Understanding small-scale astrophysical phenomena, such as star formation and feedback, is crucial for full-shape analyses. 
In this study, we measure the 3D auto-power spectrum of the $\lya$ forest and its cross-power spectrum with halos %--- rather than quasars --- 
using hydrodynamic simulations from the GADGET3-OSAKA code, which includes models for star formation and supernova feedback. 
Across five astrophysical models, we find significant deviations from the Fiducial model, with $5-10\,\%$ differences for wavenumbers $k>2\,\mpcinv$ in the $\lya$ auto-power spectrum. 
The $\lya\,\times\,$halo cross-power spectra show even larger deviations, exceeding $10\,\%$ in some cases. 
Using the fitting models of \citet{Arinyo2015} and \citet{Givans2022}, we jointly fit the $\lya$ auto- and $\lya$ $\times$ halo cross-power spectra, 
and assess the accuracy of the estimated $f\sigma_8$ parameter by comparing it with the ground truth from the simulations, while varying the maximum wavenumber $\kmax$ and minimum halo mass $M_h$. 
Our results demonstrate that the extended model of \citet{Givans2022} is highly effective in reproducing $f\sigma_8$ at $\kmax\leq3.0\,\mpcinv$ for $M_h>10^{10.5} M_\odot$, % and at $\kmax\leq5.0\,\mpcinv$ for $M_h>10^{10.5} M_\odot$. We also find that this model can reliably determine $f\sigma_8$ without being influenced by astrophysical uncertainties.
and remains robust against astrophysical uncertainties. 
\end{abstract}


%%%%%%%%%%%%%%%%%%%%%%%%%%%%%%%%%%%%%%%%%%%%%%%%%%

%%%%%%%%%%%%%%%%% BODY OF PAPER %%%%%%%%%%%%%%%%%%
%======================================================================================
%
\section{Introduction}
%
%======================================================================================
In recent decades, cosmological research has predominantly focused on the $\Lambda$CDM (Lambda Cold Dark Matter) model, spurred by the discovery of the universe's accelerated expansion. The $\Lambda$CDM model incorporates the concept of dark energy, with significant research efforts aimed at understanding its properties. The accelerated expansion is directly observed through the redshift dependence of distances and expansion rates, notably demonstrated by luminosity distances to Type Ia supernovae (SNe Ia). This phenomenon, first reported in studies by \citet{Riess1998} and \citet{Perlmutter1999}, has driven further extensive exploration.

Large-scale structure surveys, such as galaxy surveys, are powerful tools for constraining the parameters of the $\Lambda$CDM model. These surveys offer three-dimensional clustering information by measuring spectroscopic redshifts. The baryon acoustic oscillation (BAO) scale, first measured by \citet{Eisenstein2005} and \citet{Cole2005}, serves as a critical standard ruler \citep[e.g.][]{SeoEisenstein:2005}. It provides constraints on the expansion history of the universe by offering distances and expansion rates normalized to the sound horizon, observable through the BAO feature in the matter correlation function \citep[e.g.][]{Aubourg2015}.
Redshift space distortions (RSD) \citep[RSD; e.g.][]{SargentTurner1977,Kaiser1987,Hamilton1992,Cole2005,Hamilton1998}, observed as anisotropies in galaxy clustering caused by the radial components of galaxy peculiar velocities, reveal the growth rate of structure in the universe. Various redshift surveys have constrained the RSD parameter $\beta\equiv f/b$, where $f=d\ln{D}/d\ln{a}$ represents the linear growth rate, defined as the logarithmic derivative of the linear growth factor $D$ by the scale factor. 

The Lyman alpha forest (Ly$\alpha$ forest), the absorption features in the quasar's spectrum, was first noted by \citet{Lynds1971} and studied by \citet{Sargent1980} and high-resolution spectroscopy have provided rich data over the two decades \citep[e.g.][]{Weymann1981,Cowie1995,Rauch1998}. Alongside those observational endeavours, cosmological hydrodynamic simulations have been instrumental in enhancing our comprehension of the characteristics of the Ly$\alpha$ forest clouds \citep{Cen1994, Hernquist1996, Miralda-Escude1996, Croft1998,Zhang1997,Zhang1998}.

The Ly$\alpha$ forest serves as a pivotal tool for revealing the large-scale structure of the intergalactic medium (IGM), which probes cosmic expansion through the BAO feature and the growth of cosmic structures via redshift-space distortions (RSD) at higher redshifts ($z \geq 2$), surpassing the reach of previous galaxy surveys.  

Over the past decade, the Baryon Oscillation Spectroscopic Survey \citep[BOSS;][]{Dawson2013} and its extension \citep[eBOSS;][]{Dawson2016}, have been pivotal in advancing our understanding of BAO. Together, BOSS and eBOSS have amassed over 341,000 quasar spectra at redshifts $z>1.77$.
Notably, the first detections of BAO from the Ly$\alpha$ auto-correlation function and its cross-correlation with quasars were made with data from BOSS DR9 data \citep{Busca2013,Slosar2013,Kirkby2013} and DR11 \citep{Font2014}, respectively. 
Leveraging these extensive datasets, researchers successfully measured the 3D auto-correlation of the Ly$\alpha$ forest in quasars at $z>2.1$ and its 3D cross-correlation with quasar positions. This comprehensive analysis has enabled the detection of BAO around $z=2.3$, showcasing the power of increasingly larger datasets in advancing our knowledge of cosmic structures.

On the other hand, the RSD analysis using only the Ly$\alpha$ auto-correlation has not been performed, primarily due to the degeneracy of the growth rate $f$ with an unknown velocity gradient bias \citep{Seljak2012}. This problematic bias originates from the two-point correlation of the flux map, which undergoes nonlinear distortion relative to optical depth. 

Recently, \citet{Cuceu2021} proposed a joint analysis of $3\times2$-pt correlation functions ($\lya\,\times\,\lya$, $\lya\,\times$\, QSO, and QSO$\,\times\,$QSO) to measure the product of the growth rate and the amplitude of matter fluctuations, $f\sigma_8$. Note, however, that Ly$\alpha$ auto-correlation still has the degeneracy of the growth rate $f$ with an unknown bias and $f$ is measured mainly by cross-correlation and QSO auto-correlation. Additionally, they emphasized that for precise measurement of the growth rate, the modelling of the cross-correlation function must be accurate.

\citet{Givans2022} used the Sherwood suite of hydrodynamic simulations \citep{Bolton2017} and presented the measurements of the 3D power spectrum of Ly$\alpha$ auto- and of its cross-correlation with halos. They compared the fitting function approach to the perturbation theory approach as a description of the 3D power spectrum beyond linear scales. They demonstrated that the fitting function proposed by \citet{Arinyo2015} can predict a 3D power spectrum with residuals smaller than $5\%$ on scales $k<10\,\mpcinv$. Furthermore, they measured Ly$\alpha$ cross-correlation with only massive halos ($M_\mathrm{halo}>10^{11.5}M_\odot$), and it had less power than the correlation with all halos, which suggests the need to create a fitting function describing the strong loss of cross-correlation on small scales. They proposed a new model (called the `Extended model' in their paper) that can improve the fits, especially at $k\sim 1\, \mpcinv$ compared to the Kaiser model. However, we note that the L160\_N2048 model from the Sherwood simulation does not include self-shielding effect (i.e. optically thin approximation), and uses \texttt{Quick-Lya} which removes high-density gas in dark matter halos in order to avoid heavy computational load. 

The primary objective of this paper is to assess the robustness of $f\sigma_8$ measurements, with a specific focus on including small-scale correlations and the impact of varying feedback models and self-shielding effects.
We utilize outputs from the GADGET3-OSAKA cosmological smoothed particle hydrodynamics (SPH) code \citep{Aoyama2017,Shimizu2019}, conducting simulations with and without supernova (SN) feedback, along with different treatments of UV background (UVB) radiation and self-shielding corrections. These simulation sets were also analyzed in \citet{Nagamine2021}. 
By employing a variety of models within our simulation dataset and avoiding the use of the \texttt{Quick-Lya} approximation, which overlooks detailed astrophysical processes at smaller scales, we aim to elucidate the critical roles of astrophysical processes, including star formation, SN feedback, and the self-shielding effect.

As a precursor to cosmological analysis by the Subaru Prime Focus Spectrograph (PFS), we examine the three-dimensional power spectrum of $\lya$ auto- and cross-correlation with halos and perform parameter estimation by fitting those correlations. We note that we adopt only the fitting function approach, rather than the perturbation theory approach, due to our focus on particularly small scales where the density fluctuations of gas are fully non-linear and influenced by astrophysical processes. Additionally, we do not address the non-linear model of the halo auto-power spectrum, as our hydrodynamical simulations are not the best tool to discuss it.

This paper is organized as follows. In Section~\ref{subsec21}, we describe the details of our cosmological hydrodynamic simulations. The method for generating the Ly$\alpha$ forest data set and halo density distribution are presented in Section~\ref{subsec22} and \ref{subsec23}, respectively. In Section~\ref{sec:model_of_power}, we review existing models of the Ly$\alpha$ auto-power spectrum and the Ly$\alpha$-halo cross-power spectrum. Section~\ref{subsec:fitting_analysis} introduces the methods of fitting analysis and parameter estimation. In Section~\ref{sec:results}, we present the power spectrum measurements, compare them with different astrophysical models, estimate parameters, and provide the results of $f\sigma_8$ measurements through the joint fitting of the Ly$\alpha$ auto-power spectrum and $\lya\,\times$\, halo cross-power spectrum. We then discuss the validity of the fitting models in relation to the $f\sigma_8$ estimation. Finally, we summarize our findings and conclude in Section~\ref{sec:conclusions}.

%======================================================================================
%
\section{Simulations and Methods}\label{sec:methods}
%
%======================================================================================
In this section, we describe the detailed setup of our cosmological hydrodynamical simulation and the method of constructing the $\lya$ forest data.
%
%--------------------------------------------------------------------------------------
\subsection{Cosmological Hydrodynamic Simulations}\label{subsec21}
%--------------------------------------------------------------------------------------

We use the cosmological SPH code GADGET3-OSAKA \citep{Aoyama2017,Shimizu2019,Nagamine2021}, which is a revised version of GADGET-3 \citep[originally described by][as GADGET-2]{Springel2005}.
It includes treatment for star formation, SN feedback, dust formation and evolution \citep{Aoyama2017, Shimizu2019}. Our code also incorporates several significant enhancements, such as the density-independent, pressure-entropy formation of SPH \citep{Hopkins2013, Saitoh2013}, the time-step limiter \citep{Saitoh2009}, and a quintic spline kernel \citep{Morris1996}. 

Following \cite{Nagamine2021}, we employ the same simulation dataset across various physical models, as detailed in Table \ref{tab:simlist}.
Among these models, the self-shielding from the UVB is notably considered for its potential impact on the Ly$\alpha$ forest. The simulation incorporating self-shielding is referred to as the ``Shield" model, while its counterpart without self-shielding is designated as the ``Fiducial" model.

In the ``NoFB" model, SN feedback is not included to investigate the influence of the Osaka feedback model on the Ly$\alpha$ forest statistics. 
The constant-velocity galactic wind model by \citet{Springel2003} is used in the ``CW" model.   
The UVB model of \citet{FG2009} is used in the ``FG09" model instead of \citet[][hereafter HM12]{HM2012} model to assess the impacts arising from variations in the UVB model.

We use box sizes of $100\,h^{-1}$Mpc and $2\times512^3$ particles, where Mpc stands for comoving mega-parsecs. 
The adopted cosmological parameters are $(\Omega_\mathrm{m},\Omega_\mathrm{dm},\Omega_{\mathrm{b}},\sigma_8,h, n_s)=(0.3089,0.2603,0.04860,0.8150,0.6774, 0.9667)$ from \citet{Planck2016}.

\begin{figure*}
\includegraphics[width=\linewidth]{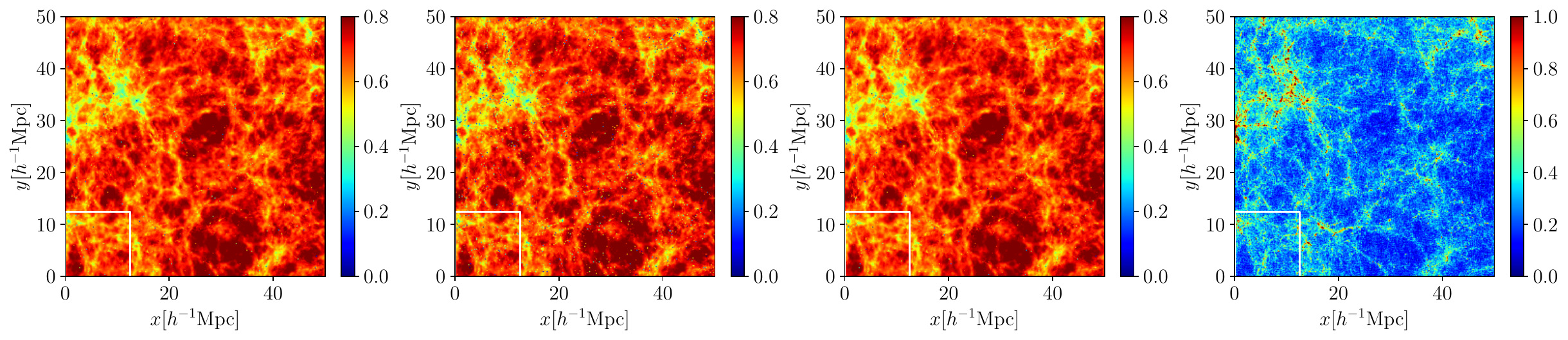}
\includegraphics[width=\linewidth]{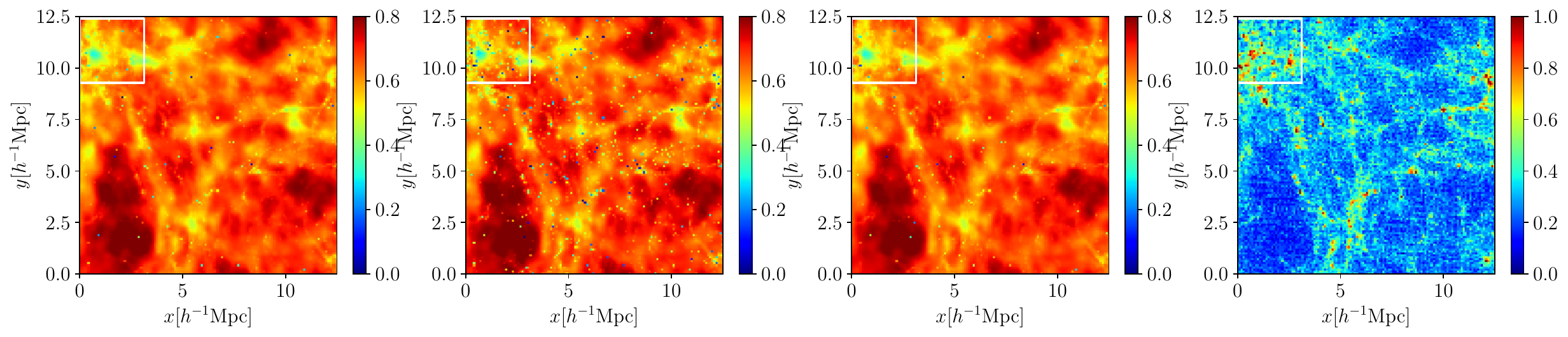}
\includegraphics[width=\linewidth]{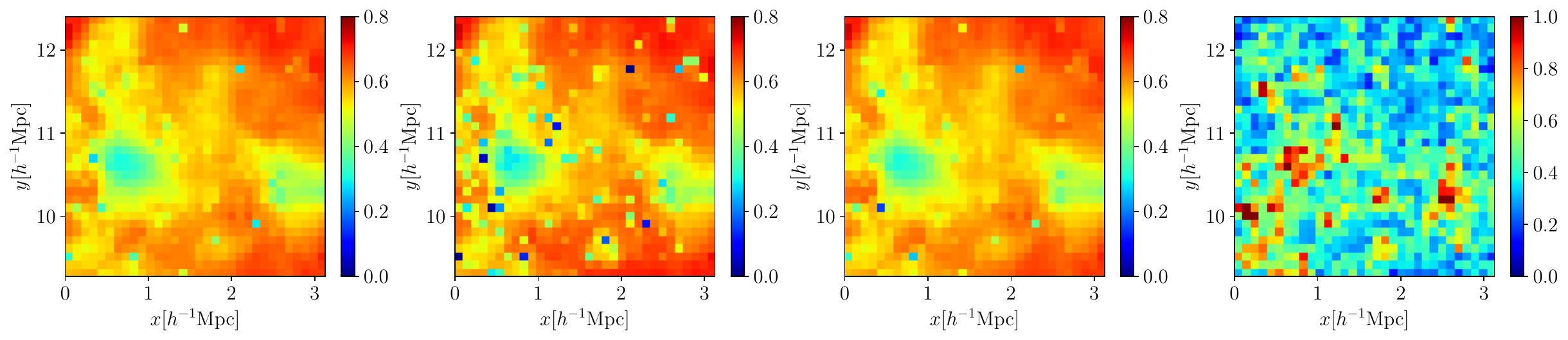}
 \caption{%25 Mpc/$h$ slice of the $\lya$ transmission $F(z)$ for fiducial, Shield and FG09 models from left to right. 
 $\lya$ transmission $F(z)$ projected over $25 \mpc$ depth along the line of sight for fiducial, Shield, and FG09 models from left to right.
 The rightmost panels show the halo distribution, where the colour code represents the normalized number density of the halo. Three panels are in redshift space, and depth is the direction of the line of sight. Top panel is the cutout of the square region of $50 \mpc$ on a side from the full simulation box. Lower panels show the closed-up map of the white square regions in the upper panels. In the bottom panels, we see some dark blue spots (i.e., $F=0$), corresponding to high HI column density systems. % caused by the Damped Lyman alpha (DLA) system.
 }
 \label{fig:2d}
\end{figure*}

%--------------------------------------------------------------------------------------
\subsection{Simulated Ly$\alpha$ forest Data Set}\label{subsec22}
%--------------------------------------------------------------------------------------

For the Ly$\alpha$ forest analysis, we follow the same methodology as in \citet{Nagamine2021}. 
We first compute the physical quantities at $i$-th pixel at three dimensional coordinate $\boldsymbol{x}_i$ as 
\begin{align}
    A_\mathrm{pixel}(\boldsymbol{x}_i) = \sum_n \frac{m_n}{\rho_n}A_n{W}(r_{in},h_n),
\end{align}
where sum models over all particle and $A_n$ represents physical quantities assigned to the $n$-th particle, such as hydrogen density $n_\mathrm{HI}$, velocity $v$, and temperature $T$,
$m_n$ is the gas particle mass, $\rho_n$ is the gas density, $h_n$ is the smoothing length of the particle and $r_{in}=|\boldsymbol{x}_i-\boldsymbol{q}_n|$ is the distance between the pixel $\boldsymbol{x}_i$ and particle position $\boldsymbol{q}_n$. 
For the kernel function, we use the following cubic spline kernel \citep{Monaghan1985}:
\begin{equation}
    W^\mathrm{SPH}(r,h)=\frac{8}{\pi h^3}
    \begin{cases}
    1-\frac{3}{2}\left(\frac{r}{h/2}\right)^2+\frac{3}{4}\left(\frac{r}{h/2}\right)^3 & 0 < r < h/2\\
    \frac{1}{4}\left(2-\frac{r}{h/2}\right)^3 & h/2 <r<h\\
    0 & h<r
    \end{cases}
    .
\end{equation}
So far, we use the single index of $i$ for uniquely identifying the pixel position for simplicity, but let us rewrite the coordinate as $\boldsymbol{x}=(x_i, y_j, z_k)$. We assume, without losing the generality, that the $z$ is the line of sight direction (LoS).
Then, we use the $A_\mathrm{pixel}(\boldsymbol{x}_i)$ values to compute the $\lya$ optical depth $\tau$ at any LoS position of $z$ but on the grid $(x_i,y_j)$,
\begin{align}
    \label{eq:tau}
    \tau_{ij}(z) = \frac{\pi{e}^2}{m_e{c}}f\sum_k\phi(z,z_k | T_{ijk},v_{ijk})n_\mathrm{HI}(x_i,y_j,z_k)dl,
\end{align}
where $f$ and $n_\mathrm{HI}$ are absorption oscillator strength, and HI number density. The pixel length, $dl$, is set to a constant, $L_\mathrm{box}/N_\mathrm{pixel}\sim0.1\mpc$. The function $\phi$ is the Voigt profile which represents the line profile of the $\lya$ absorption. We can obtain it by averaging the Lorentzian profile for a single atom over a collection of atoms with a Maxwellian distribution of thermal velocities. The gas temperature $T$ characterizes the thermal broadening of spectral lines due to the Doppler effect, and the peculiar velocity $v$ distorted the spatial distribution of absorption lines due to the Doppler shift in addition to the redshift caused by the cosmological expansion. Here, we utilize the fitting formula of \citet{Tasitsiomi2006} which is accurate for temperatures $T>2\mathrm{K}$, rather than direct integration.

The observed $\lya$ transmitted flux can be written as $F(z)=\exp(-\tau(z))$. Finally, we adjust the mean transmitted flux fraction $\bar{F}(z)$ corresponding to the intensity of the ionizing background. The simulated transmitted flux is normalized to the observed value of $\bar{F}(z)$. Here, we determine $\bar{F}$ so that the effective optical depth $\tau_\mathrm{eff}(z)=-\ln\langle\bar{F}(z)\rangle$ is identical to that of \citet{Becker2013} which is calculated by 
\begin{align}
    \tau_{\mathrm{eff}}(z)=\tau_0\left(\frac{1+z}{1+z_0}\right)^\beta+C,
\end{align}
and we use their best-fitting parameters $[\tau_0, \beta, C]=[0.751, 2.90,-0.132]$ for $z_0=3.5$. They ensured that this analytic fit closely matches the binned $\tau_\mathrm{eff}(z)$ at all redshifts, while scaling C to literature values around $z\sim2$.
As we put the subscription of $ij$ in equation (\ref{eq:tau}), $F$ can be generally written as $F(\boldsymbol{x})$. 

Left 3 panels in Fig.~\ref{fig:2d} show the 2D slice of the transmitted flux fraction $F(x)=e^{-\tau(x)}$  at $z=3$ averaged over a depth of $25\ h^{-1}$Mpc. We see some differences between three models on color maps. The Shield model in the second left has clumping structure of low $F$ marked as blue regions, especially in the bottom panels.

\begin{threeparttable}
 \caption{Physical models employed in Osaka20 simulations}
 \label{tab:simlist}
 \centering
  \begin{tabular}{lc}
  \hline
  \hline
  Model & Notes\\
  \hline
  Fiducial & No self-shielding\\
  Shield & With self-shielding\\
  NoFB & No SN feedback\\
  CW & Constant-velocity galactic wind model\tnote{a}\\
  FG09 & UVB model of FG09\tnote{b}\\
   \hline
\end{tabular}
 \begin{tablenotes}
 \item[a] \citet{Springel2003}
 \item[b] \citet{FG2009}
 \end{tablenotes}
\end{threeparttable}

%--------------------------------------------------------------------------------------
\subsection{Halo density distribution}\label{subsec23}
%--------------------------------------------------------------------------------------

Our simulations include a halo catalog constructed with a friends-of-friends algorithm. 
We measure the halo mass function in our simulation and find that it agrees with the theoretical prediction \citep{Sheth99} for halo masses $10^{10} \lesssim \mh/M_\odot \lesssim 10^{13}$.
Halo samples are assigned on the grid using a nearest-grid-point interpolation. To measure the $\lya$ $\times$ halo cross-spectrum, we create a $1024^3$ density field of halo same as the 3D distribution of transmitted flux $F$. 

The halos in this paper are assumed to be quasars and galaxies in actual observations. The halo distribution should be given in redshift space with a view to measuring the structural growth rate by RSD analysis. Positions of each halo in real space $(x_r,y_r,z_r)$ are shifted to in redshift space $(x_s, y_s, z_s)$, assuming $z_r$ and $z_s$ are the LoS position, according to the following formula:
\begin{align}
(x_s, y_s, z_s) = \left(x_r,y_r,z_r+\frac{v_\mathrm{pec}}{aH}\right),
\end{align}
where $a$ is the scale factor, $H$ is the Hubble parameter and $v_\mathrm{pec}$ is the peculiar velocity in LoS direction. The rightmost panels in Fig.~\ref{fig:2d} show a 2D slice of the halo distribution. Comparing this with the $\lya$ flux in the left three panels, we observe an anti-correlation between halo number density and Ly$\alpha$ flux. Consequently, the $\lya\,\times\,$halo cross-power spectrum is negative on most wavenumber scales.

%======================================================================================
%
\section{Model of the power spectrum}\label{sec:model_of_power}
%
%======================================================================================
%
%--------------------------------------------------------------------------------------
\subsection{Ly$\alpha$ auto power spectrum}
%--------------------------------------------------------------------------------------
The transmitted flux $F$ is the ratio of the observed flux to the continuum flux emitted by the background source. The fluctuation of the transmitted flux $F$ is defined as
\begin{align}
    \delta_F=\frac{F}{\bar{F}(z)}-1 ,
\end{align}
where $\bar{F}$ is the mean value of transmitted flux for the same redshift.

The linear expression of the $\delta_F$ field in redshift space is 
\begin{align}
    \delta_F = b_{F} \,\delta_\mathrm{m} + b_{\eta}\,\eta ,
\end{align}
where $\delta_\mathrm{m}$ is the matter density fluctuation, and 
\begin{align}
\label{eq:eta}
    \eta=-\frac{1}{aH}\frac{\partial{v}_\mathrm{los}}{\partial{x_\mathrm{los}}},
\end{align}
is the dimensionless gradient of the peculiar velocity $v_\mathrm{los}$ along the LOS. Here, $ x_\mathrm{los}$ is the comoving coordinate, $a$ is the scale factor, and $H$ is the Hubble parameter. In Fourier space, equation (\ref{eq:eta}) is transformed as $\eta=f\mu^2\delta_\mathrm{m}$, where $f$ is the growth rate and $\mu$ is the cosine of the angle between the Fourier mode $\bf{k}$ and the LOS. The linear power spectrum of $\delta_F$ in redshift space can be written as 
\begin{align}
    \pf=b^2_{F}(1+\beta_F\mu^2)^2\pl,
\end{align}
where $\beta_F=b_\eta{f}/b_F$ is the RSD parameter and $\pl$ is the linear matter power spectrum. 

Multiplicative corrections to the linear model are useful to express non-linearity on small scales as below:
\begin{align}
\label{eq:autopower}
    \pf=b^2_{F}(1+\beta_F\mu^2)^2\pl\dnl, 
\end{align}
and \citet{McDonald2003} presented the parameterized correction term $\dnl$ with eight free parameters that could fit the flux power spectrum down to $k=10\,\mpcinv$. \citet{Arinyo2015} also introduced their fitting function with six parameters as follows:
\footnotesize
\begin{align}
\label{eq:arinyo}
    \dnl=\exp\left\{[q_1\Delta^2(k)+q_2\Delta^4(k)]\left[1-\left(\frac{k}{k_v}\right)^{a_v}\mu^{b_v}\right]-\left(\frac{k}{k_p}\right)^2\right\},
\end{align}
\normalsize
where $\Delta^2(k)=k^3P_\mathrm{L}(k)/2\pi^2$ is the dimensionless linear matter power spectrum. We will call equations\,(\ref{eq:autopower}) and (\ref{eq:arinyo}) as the {\it Arinyo} model. Here, \citet{Arinyo2015} and \citet{Givans2022} mentioned that the $q_2$ parameter in this model was unnecessary. We also find that the best-fit value for $q_2$ is very close to zero. Therefore, we ignore this second term, and the number of free parameters is reduced to five. 

%--------------------------------------------------------------------------------------
\subsection{Ly$\alpha\,\times\,$halo cross power spectrum}
%--------------------------------------------------------------------------------------

Before considering the model of the cross-power spectrum, we introduce that of the halo auto power spectrum. In a linear regime, the standard model of the halo auto power spectrum in redshift space, which was presented by \citet{Kaiser1987} is
\begin{align}
\label{eq:pk_halo_lin}
    P_h(k,\mu)=(b_h+f\mu^2)^2P_\mathrm{L}(k),
\end{align}
where $b_h$ is the linear halo bias which serves as a proportionality factor connecting fluctuations in halo number density to fluctuations in dark matter mass density, and it proves to be a reliable approximation on large scales.

The model of the Ly$\alpha$ $\times$ halo cross-power spectrum was first developed by \citet{Givans2022}. On large scales, $\lya$ forest and halo auto-power spectrum follow the Kaiser model (eq. (\ref{eq:autopower}) and (\ref{eq:pk_halo_lin})), and $\lya$ $\times$ halo cross-power spectrum is expected to be expressed as the geometric mean of the Kaiser models for both tracers:
\begin{align}
\label{eq:pk_cross_lin}
    P_{\times,{\rm L}}(k,\mu)&=\sqrt{\pf P_h(k,\mu)}\\
    &=b_F(1+\beta_F\mu^2)(b_h+f\mu^2)P_\mathrm{L}(k).
\end{align}
Similarly, the flux component of the non-linear correction can be modelled as the square root of equation (\ref{eq:arinyo}) from the {\it Arinyo} model:
\begin{align}
\label{eq:pk_cross_dnl}
    P_{\times}(k,\mu)=b_F(1+\beta_F\mu^2)(b_h+f\mu^2)P_\mathrm{L}(k)\sqrt{D_\mathrm{NL}(k,\mu)}.
\end{align}
They referred to this equation as the \mini model, and we use the same name. Then, \citet{Givans2022} proposed that massive halos hosting quasars are less correlated with the Ly$\alpha$ forest, making it challenging to fit the cross-correlation on small scales. As a result, they developed a more complex model:

\begin{align}
\label{eq:crosspower}
P_\times(k,\mu)=b_F(1+\beta_F\mu^2)(b_h+f\mu^2)P_\mathrm{L}(k)\sqrt{\dnl}\dm,
\end{align}
\begin{align}
\label{eq:givans}
\dm=\exp\left[(\alpha+\gamma\mu^2)\Delta^2(k)-(k\mu\nu)^4\right],
\end{align}
where $\alpha$, $\gamma$, $\nu$ are the free parameters. Equations\,(\ref{eq:crosspower}) and (\ref{eq:givans}) are used in our main analysis in Section~\ref{subsec:fs8constraints} and are referred to as the {\it Givans} model.

On the other hand, observations could measure $\lya$ cross-correlation with quasars, etc., rather than halos. Quasar non-linear velocities, commonly known as the 'Fingers of God' (FoG) effect, is an important non-linear effect for the quasar auto- and cross-correlation with $\lya$ forest. Following  \citet{Percival_White_2009}, we also use
\begin{align}
\label{eq:Bourboux}
    P_\times(k,\mu)=b_F(1+\beta_F\mu^2)(b_h+f\mu^2)P_\mathrm{L}(k)\sqrt{\fog},
\end{align}
\begin{align}
\label{eq:fog}
    \fog=\frac{1}{1+(k\mu\sigma_v)^2},
\end{align}
where $\sigma_v$ is a free parameter representing the rms velocity dispersion. This model has been used for quasar redshift surveys and their forecast \citep[e.g.][]{Bourboux2020, Cuceu2021, Adame2024}, and it also accounts for statistical quasar redshift errors. We refer to this equation as the {\it FoG} model and use it for the fitting of the $\lya$ $\times$ halo cross-power spectrum to estimate $f\sigma_8$.

To compute the linear matter spectrum $P_\mathrm{L}(k)$, we use CAMB code with input parameters fixed to their values adopted in our simulation. When performing a fitting analysis, free parameters for the linear RSD are $b_F\sigma_8$, $b_h\sigma_8$ and $f\sigma_8$ because $\pl\propto\sigma_8^2$. 

In Section~\ref{subsec:compare_dnl_dnldm}, we present the results of the $f\sigma_8$ estimation using equations (\ref{eq:pk_cross_dnl}), (\ref{eq:crosspower}) and (\ref{eq:Bourboux}), comparing the outcomes from each to evaluate the validity of these fitting models. We summarize these fitting models in Table~\ref{tab:fitting_models}.

\begin{table}
 \caption{Fitting models of $\lya$ $\times$ halo cross-power spectrum. The rightmost column shows the number of free parameters in joint fitting of $\pf$ and $\pc$.}
 \label{tab:fitting_models}
 \centering
  \begin{tabular}{lccc}
  \hline
  \hline
  Model & Equation & Non-linear term & Number of Parameter \\
  \hline
  \mini & (\ref{eq:pk_cross_dnl}) & $\sqrt{\dnl}$ &  8\\
  
   {\it Givans} & (\ref{eq:crosspower}) & $\sqrt{\dnl}\dm$ & 11 \\
   
   {\it FoG} & (\ref{eq:Bourboux}) & $\sqrt{\fog}$ &  9\\
  \hline
  
\end{tabular}
\end{table}
%--------------------------------------------------------------------------------------
%
\begin{figure*}
 \begin{tabular}{c|c}
\includegraphics[width=0.45\linewidth]{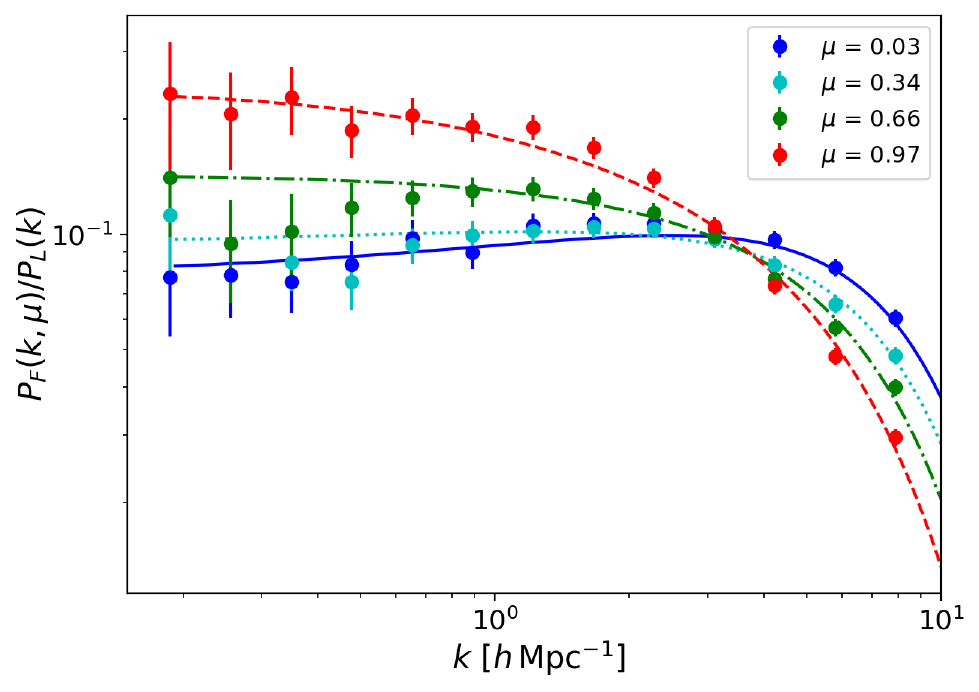} &
\includegraphics[width=0.45\linewidth]{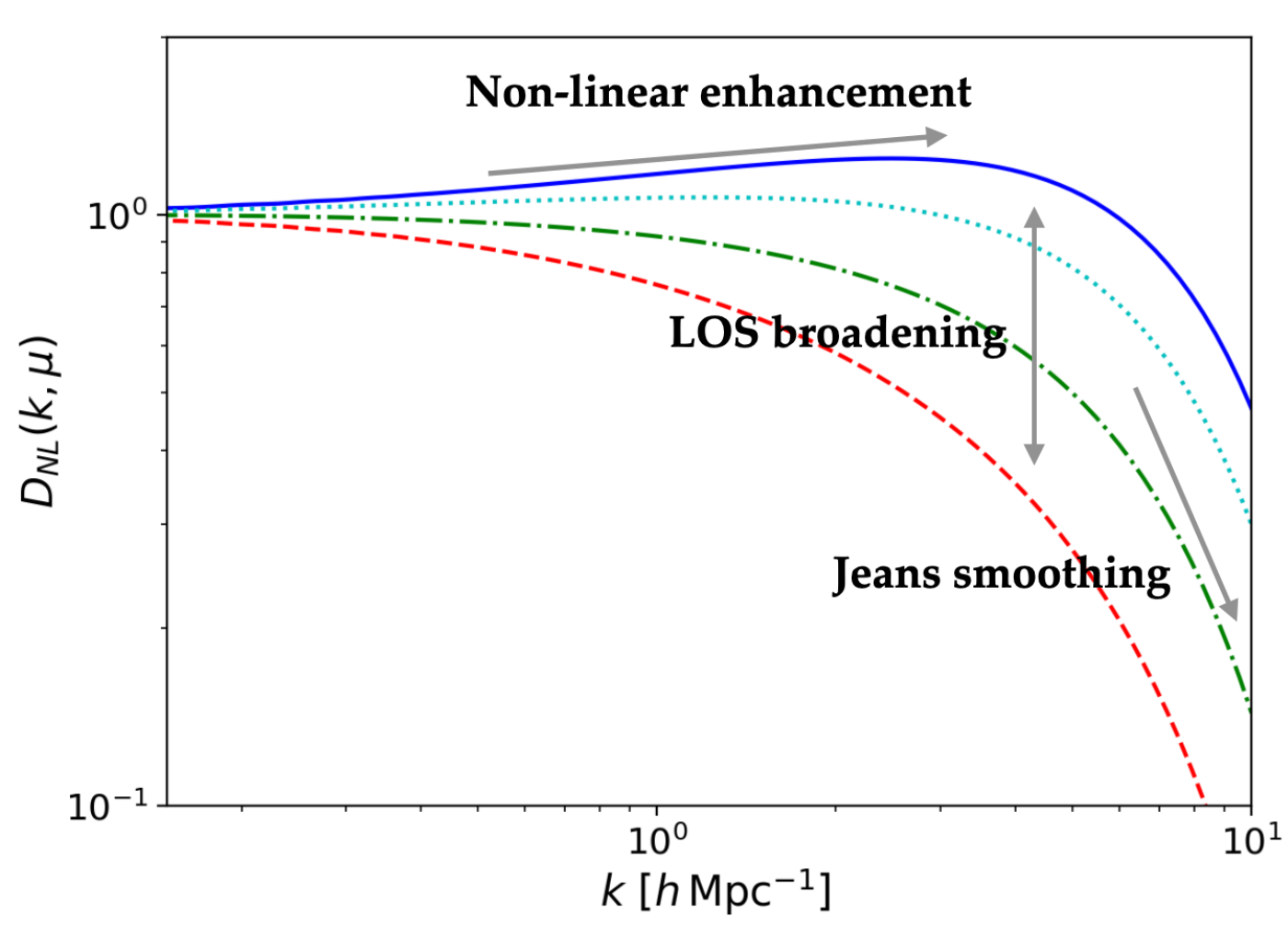}\\
 \end{tabular}
 \caption{\textit{Left} : Measured and best-fitting power spectrum. Points and errorbars are measured Ly$\alpha$ auto power spectra using datasets of Fiducial model, normalized by the linear power spectrum. Lines are the best-fitting models using the Arinyo model (equations\,(\ref{eq:autopower}) and (\ref{eq:arinyo})) for $k<10\,\mpcinv$, where the maximum $k$-bin represented by $k=7.90\,\mpcinv$ covered from $6.75\,\mpcinv$ to $9.22\,\mpcinv$.  The \textit{right} panel schematically describes the physical dependencies of the non-linear correction term $D_{\rm NL}$ on $k$ and $\mu$. 
}
 \label{fig:autopower}
\end{figure*}

\subsection{Fitting Analysis}
\label{subsec:fitting_analysis}
%
%--------------------------------------------------------------------------------------
We fit models by Markov Chain Monte-Carlo (MCMC) sampling with \texttt{emcee} package \citep{Foreman2013} to minimize the pseudo-$\chi^2$ function:
\begin{align}
    \chi^2(\bm{\theta})=\sum_{i,j} \frac{[P^\mathrm{sim}(k_i,\mu_j)-P^\mathrm{model}(k_i,\mu_j|\bm{\theta})]^2}{\sigma_{ij}^2},
\end{align}
where $P^\mathrm{sim}(k_i,\mu_j)$ is the measured power spectrum for the $i-$th $k$ and $j-$th $\mu$ bin, $P^\mathrm{model}(k=k_i,\mu=\mu_j)$ is the model power spectrum given in Section~\ref{sec:model_of_power} and $\sigma_{ij}$ is the statistical uncertainty on $P^\mathrm{sim}(k_i,\mu_j)$. In the following, however, $P(k_i,\mu_j)$ will be written simply as $P_i$.We perform two types of fitting analysis in our work: only $\lya$ auto-power spectrum and joint fitting of 2$\times$2-pt correlation ($\lya\,\times\,\lya$ and $\lya\,\times\,$halo). 
In the former, we minimize the following pseudo-$\chi^2$ function:
\begin{align}
    \chi^2_F(\bm{\theta}_F)=\sum_i \frac{\left[P^\mathrm{sim}_{F,i}-P_{F,i}^\mathrm{model}(\bm{\theta}_F)\right]^2}{\sigma_{F,i}^2},
\end{align}
where the model of $\lya$ auto-power spectrum $P_F^\mathrm{model}$ is given by the equation\,(\ref{eq:autopower}) and (\ref{eq:arinyo}) with seven free parameters
\begin{align}
\label{eq:params_auto}
    \bm{\theta}_F = \{b_{F}, \beta_F, q_1, k_p, k_v^{a_v}, a_v, b_v\},
\end{align}
 using wavenumbers $k_i$ up to $\kmax=10\,\mpcinv$. 
 
 In the latter, we minimize the following pseudo-$\chi^2$ function:
\begin{align}
    \chi^2_\mathrm{joint}(\bm{\theta}_\times)=\chi^2_F(\bm{\theta}'_F)+\chi^2_{\times}(\bm{\theta}_\times),
\end{align}
where
\begin{align}
    \chi^2_\times(\bm{\theta}_\times)=\sum_i \frac{\left[P^\mathrm{sim}_{\times,i}-P_{\times,i}^\mathrm{model}(\bm{\theta}_\times)\right]^2}{\sigma^2_{\times,i}},
\end{align}
using $k_i$ up to $\kmax=1, 3, 5\,\mpcinv$. The models of cross power spectrum $P_\times^\mathrm{model}$ are given by the equation\,(\ref{eq:pk_cross_dnl})- (\ref{eq:fog}) with free parameters
\begin{align}
\label{eq:params_cross}
    \bm{\theta}'_F &= \{b_{F}, \beta_F, q_1, k_v^{a_v}, a_v, b_v\}\\
    \bm{\theta}_\times &= \{\bm{\theta}'_F, \{b_h\sigma_8, f\sigma_8\}, \bm{\theta}_{\rm \times,NL} \}
\end{align} 
where $\bm{\theta}_{\rm \times,NL}$ is the additional parameter in non-linear correction; \mini model has no $\bm{\theta}_{\rm \times,NL}$, {\it Givans} model has $\bm{\theta}_{\rm \times,NL}=\{\alpha, \gamma, \nu\}$ and {\it FoG} model has $\bm{\theta}_{\rm \times,NL}=\{\sigma_v\}$. We note that other fundamental cosmological parameters related to $\pl$ are fixed. When we perform the joint fitting, the parameters $\bm{\theta}'_F$ are shared with $\bm{\theta}_\times$ and we remove the $k_p$ parameter because we find the best-fit value for $k_p$ diverges to infinity on the scale of $k\leq5\,\mpcinv$.

In this work, the  uncertainty $\sigma_i$ is defined as the inverse of the measurement uncertainty of the power spectrum:
\begin{align}
\label{eq:error}
\sigma_{F,i}=P_{F,i}^\mathrm{sim}\left[{N_i}^{-1/2}+\epsilon\right].
\end{align}
When the density field is a Gaussian random field, the variance of the power spectrum can be estimated by the number $N_i$ of Fourier modes contributing to each bin \citep{Jeong2009}. 
However, the high-$k$ data points with exceptionally small error bars (thanks to high $N_i=N(k_i,\mu_i)$) will dominate in the standard fit as elaborated in  \citet{McDonald2003,Arinyo2015}. Following these earlier works, our approach involves adjusting the standard weighting system, incorporating a noise floor $\epsilon=0.05$ . 

Next, we estimate the measurements uncertainty of $\lya\,\times\,$halo cross-power spectrum $\sigma_{\times,i}$, following the appendix B in \citet{Font2014}.
First, we define the weight of the halo auto-power spectrum as 
\begin{align}
\sigma_{h,i}=\left[P^\mathrm{sim}_{h,i}+n_h^{-1}\right]{N_i}^{-1/2},
\end{align}
where $n_h$ is the number density of the halo. The statistical uncertainty of ${P_{\times,i}^\mathrm{sim}}$ is derived using ${P_{\times,i}^\mathrm{sim}}$ itself, $\sigma_{F,i}$ and $\sigma_{h,i}$ as
\begin{align}
\label{eq:error_cross}
\sigma_{\times,i}=\left[\left(P_{\times,i}^\mathrm{sim}\right)^2{N_i}^{-1}+\sigma_{F,i}\sigma_{h,i}\right]^{1/2}.
\end{align}
We note that these weights are merely estimators, and it is preferable to measure the covariance directly from the simulated data. However, measuring the covariance matrix will require a larger simulation box, which will be addressed in a future study.

%======================================================================================
%

\section{Results and Discussions}\label{sec:results}
%
%======================================================================================
%
%--------------------------------------------------------------------------------------
\subsection{Ly$\alpha$ 3D Power Spectrum}\label{sec:autopower}
%--------------------------------------------------------------------------------------
This section shows the $\lya$ auto power spectrum measured from out simulation set with the best-fitting models. We compute the best-fitting parameters for different astrophysical models and discuss how large are the effects of the physical models on the $\lya$ power spectrum.

\subsubsection{Power spectrum measurements for fiducial model}

We assume that the background QSOs are uniformly distributed at the position of the regular grid and thus $1024^2$ QSOs in total in the simulation box.
The power spectrum is measured on the regular $1024^3$ grid of Fourier space grid where $k$ is logarithmically uniformly sampled in 20 bins and cosine angle from 0 to 1 into 16 bins, assuming the symmetry. The range of wavenumbers spans from $k_0 = 2\pi/L = 0.0628\, h\mathrm{Mpc}^{-1}$ to the Nyquist frequency, $32.2\,h\mathrm{Mpc}^{-1}$. 
We present the measured anisotropic power spectrum of Ly$\alpha$ transmitted flux at $z = 3.0$ in Fig.~\ref{fig:autopower}, where we show only for four $\mu$ bins and error bars are computed from equation (\ref{eq:error}).
Lines are the best-fitting model using equations\,(\ref{eq:autopower}) and (\ref{eq:arinyo}).
It is noticeable that, on large scales, the power along the LOS (red) is consistently enhanced by a constant factor in comparison to the transverse power (blue) due to the linear Kaiser effect. 
At smaller scales, the value of $\dnl$ becomes crucial. We observe power enhancement for small $\mu$ at $0.5 \mpcinv<k<3 \mpcinv$. This enhancement can be captured by parameters $q_1$ and $q_2$ in equation\,(\ref{eq:arinyo}), representing a $\mu$-independent effect originating from gravitational growth. However, this enhancement only appears at small $\mu$, because it is highly suppressed by LOS broadening for larger $\mu$. Additionally, we see power suppression along the LOS due to the non-linear peculiar velocities and the thermal broadening, captured by the parameters $k_v$, $a_v$ and $b_v$. The overall suppression on $k\lesssim 8\,\mpcinv$, independent of $\mu$, is interpreted as the smoothing effect by gas pressure, characterized by the parameter $k_p$.

\begin{figure*}
 \begin{tabular}{c|c}
\includegraphics[width=0.45\linewidth]{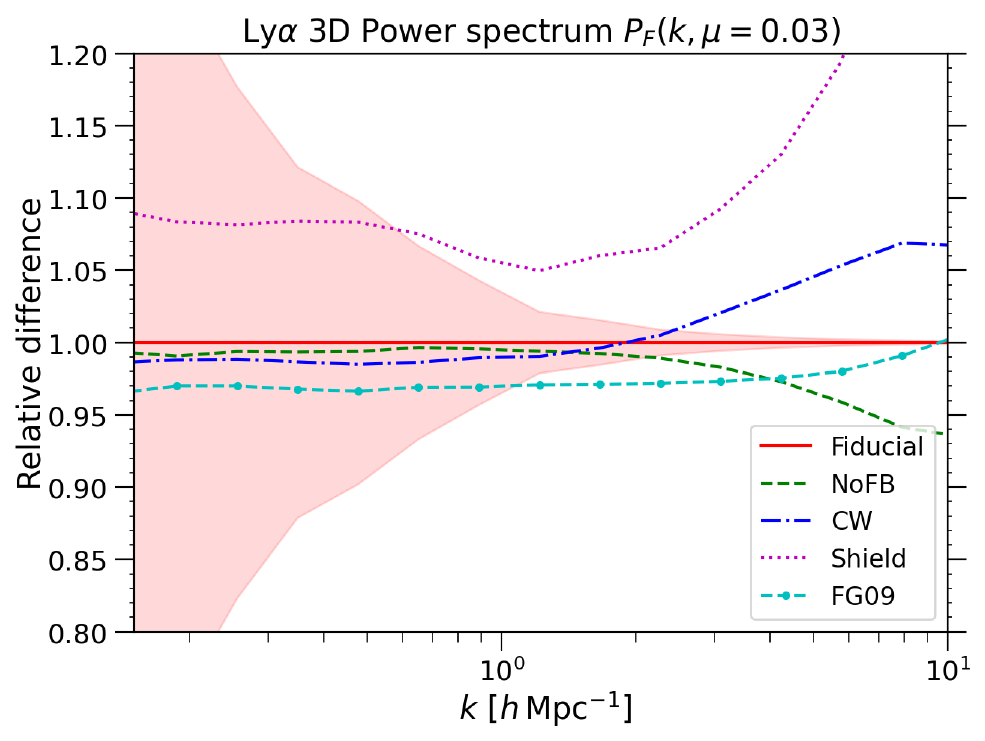}&
 \includegraphics[width=0.45\linewidth]{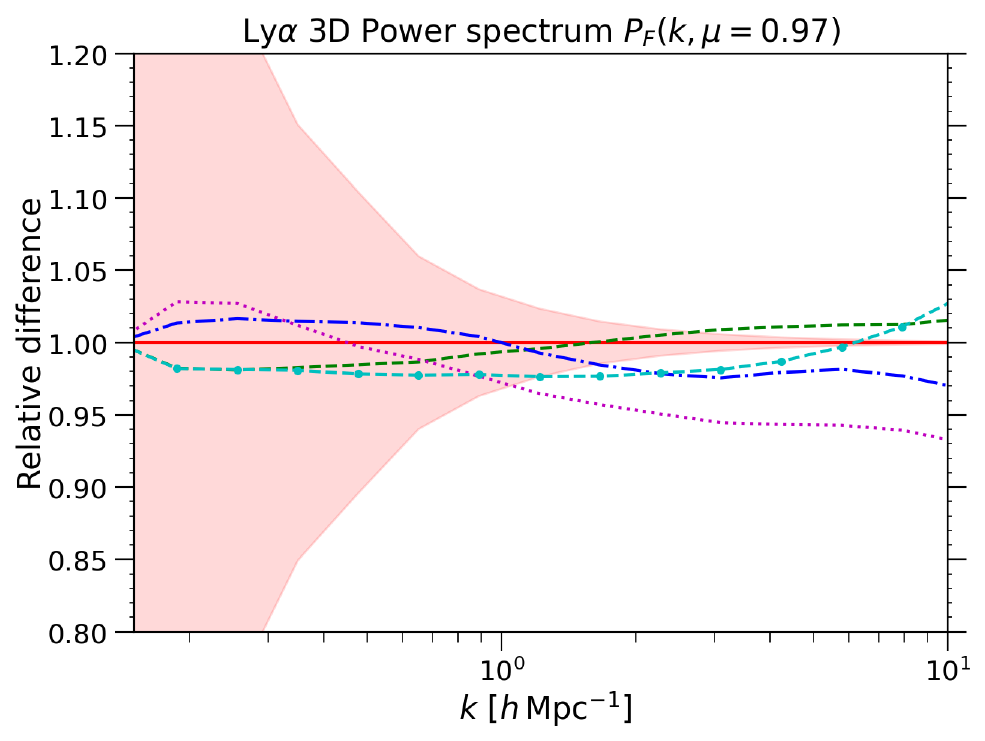}\\
 \end{tabular}
 \caption{Comparison of the Ly$\alpha$ auto power spectrum across different simulation models given in Table~\ref{tab:simlist}. The relative difference is obtained by dividing the power spectrum of each model by that of the Fiducial model, with the pink shaded areas representing the $1\sigma$ error margins of the Fiducial model calculated according to equation (\ref{eq:error}). The left panel depicts the relative difference in the flux power spectrum $P_F(k,\mu=0.03)$ compared to the Fiducial model and the right panel illustrates the relative difference in $P_F(k,\mu=0.97)$, providing insights into model performance at distinct angular directions.}
 \label{fig:comp_auto}
\end{figure*}

\begin{figure*}
\includegraphics[width=\linewidth]{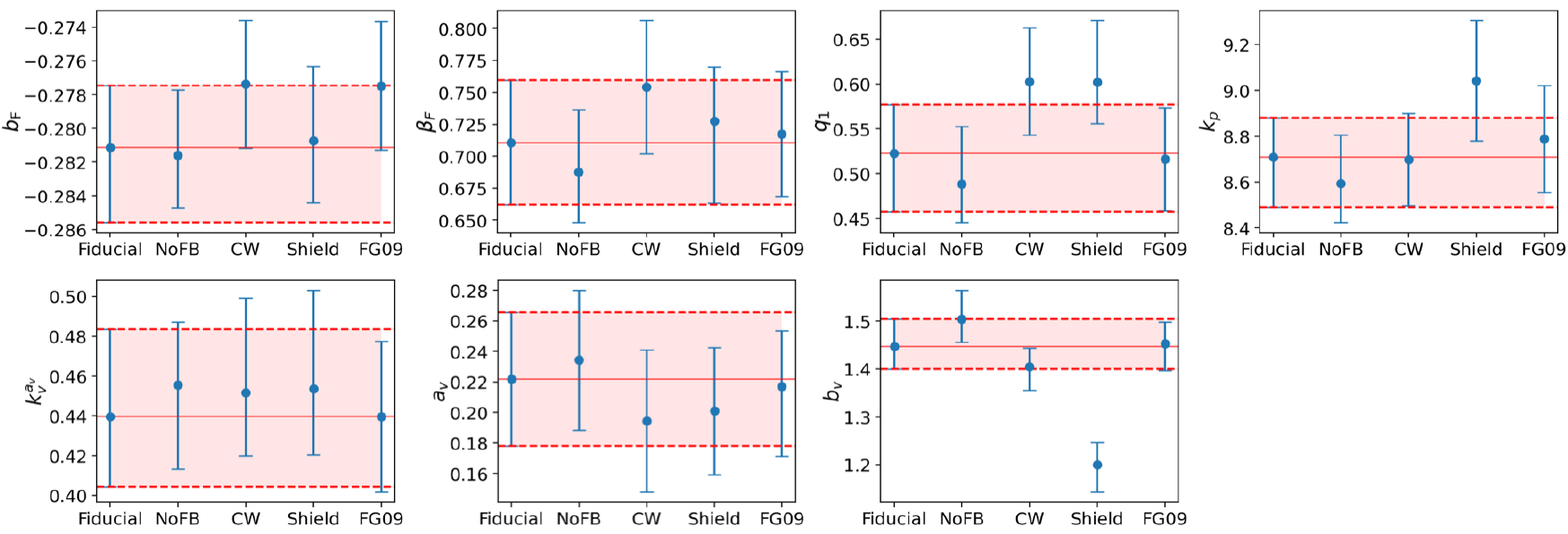}
 \caption{Fitting results of the nuisance parameters with 1$\sigma$ errors for five different astrophysical models given in Table\,\ref{tab:simlist}. 
Red shaded region is the 1-sigma region of the fiducial model. The parameter values are almost consistent with each other within the error bar, except for the $b_\nu$ in the "Shield" model.
 }
 \label{fig:compare}
\end{figure*}
\subsubsection{Comparison with different astrophysical models}

Fig.~\ref{fig:comp_auto} compares the $\lya$ auto power spectrum measured by the simulations with different astrophysical models against the Fiducial model. 
The statistical errors are depicted in shaded regions, calculated according to equation (\ref{eq:error}) in the limit of $\epsilon=0$.
In this figure, we explore two extreme cases where $\mu=0.03$ (left panel) and $\mu=0.97$ (right panel).
The left panel corresponds to the direction perpendicular to the line of sight and is considered to trace the density fluctuations of the IGM directly. The right panel, on the other hand, reflects LoS broadening due to gas temperature and local velocity, in addition to density fluctuations.

In the left panel, the NoFB, CW, and FG09 models are almost consistent with the Fiducial model within $5\%$ at $k<5\,\mpcinv$ with slight discrepancies for NoFB and CW models on smaller scales. 
Regarding the small-scale behavior of each model, the NoFB and FG09 models exhibit smaller amplitudes, whereas the CW and Shield models show larger amplitudes compared to the Fiducial model.
Notably, the Shield model exhibits significantly enhanced power relative to the Fiducial model. This discrepancy can be attributed to the more pronounced clumping structure of HI gas resulting from the self-shielding effects, which prevent the penetration of the ionizing photons, particularly at higher-density regions such as the galactic centre. 

In the right panel, all models except the Shield model show the deviation from the Fiducial model within 5\% at $k <10\,\mpcinv$.
We see that the power spectrum of the CW model is smaller than that of the Fiducial model at $k\gtrsim1\mpcinv$. This suggests that the suppression effect due to LoS broadening appears stronger than the Fiducial model, which can be explained by the strength of the heating effect.
In the CW model, the gas is ejected out of the galactic potential more efficiently compared to other models, leading to some overheating of the HI gas as documented by \citet{Choi2011}. The Shield model also shows a smaller amplitude than the Fiducial model at $k\gtrsim1\mpcinv$, whose interpretation is unclear. In the NoFB model, the stronger power compared to the Fiducial model can be explained by the less heating of the HI gas due to the absence of supernova feedback. However, the difference is only a few percent.
In this context, $P_F(k,\mu=0.97)$ represents the correlation in redshift space along the LOS, aiming to reflect properties akin to those found in the 1D power spectrum along the LOS. Indeed, this observation aligns with trends reported in $P_{1D}(k)$ measurements by \citet[][Fig.\,4d]{Nagamine2021}, highlighting a consistent pattern across different analyses.

\begin{figure*}
 \begin{tabular}{c|c}
 \begin{minipage}[t]{0.45\hsize}
 \includegraphics[width=\columnwidth]{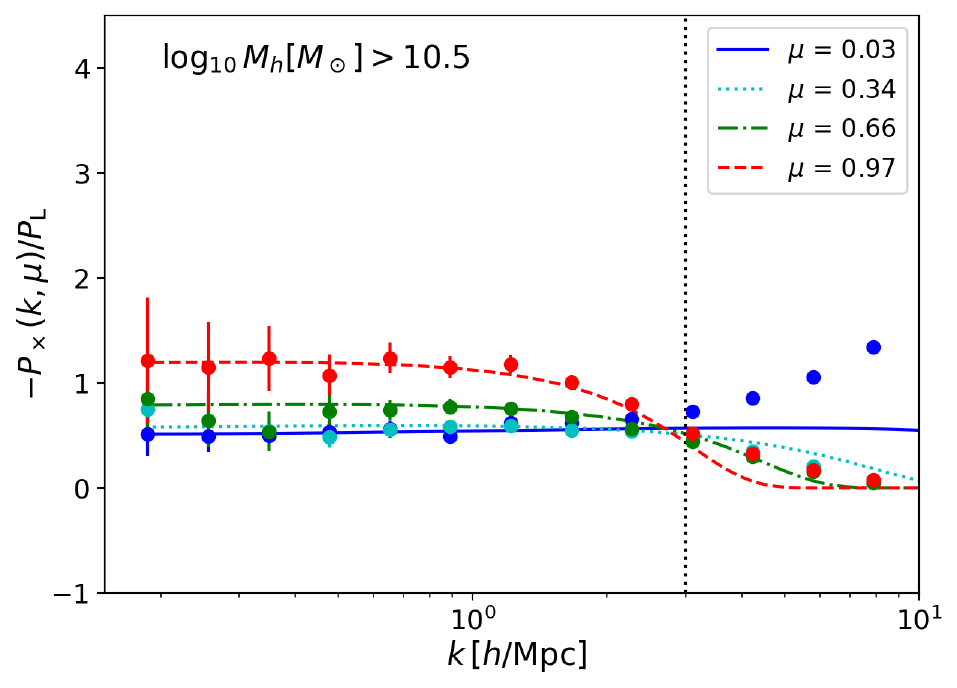}
 \end{minipage}
 \begin{minipage}[t]{0.45\hsize}
 \includegraphics[width=\columnwidth]{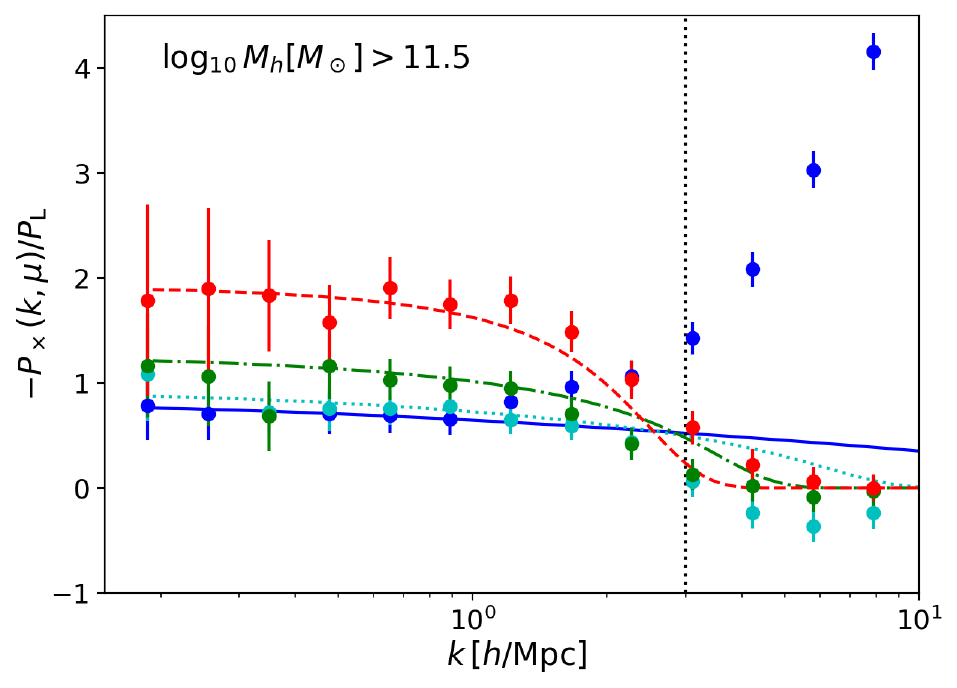}
 \end{minipage}
 \end{tabular}
 \caption{Points and errorbars are measured power spectra of $\lya\,$ from Fiducial model, cross-correlated with a halo of limiting mass $10^{10.5}M_\odot$ (left) and $10^{11.5}M_\odot$ (right). Lines are the best-fitting models from the joint fitting of $\lya$ auto- and cross-power spectrum with halo using scales up to $\kmax=3\,\mpcinv$ (vertical dotted line). 
}
 \label{fig:crosspower}
\end{figure*}

\begin{figure*}
 \begin{tabular}{c|c}
 \begin{minipage}[t]{0.45\hsize}
 \includegraphics[width=\columnwidth]{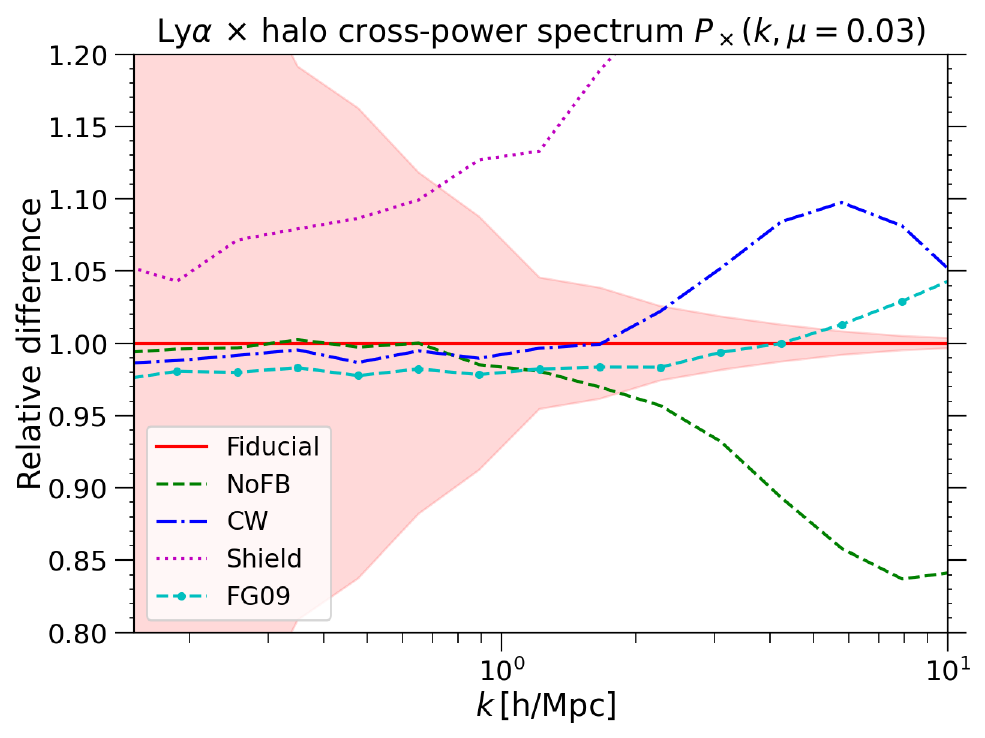}
 \end{minipage}
 \begin{minipage}[t]{0.45\hsize}
 \includegraphics[width=\columnwidth]{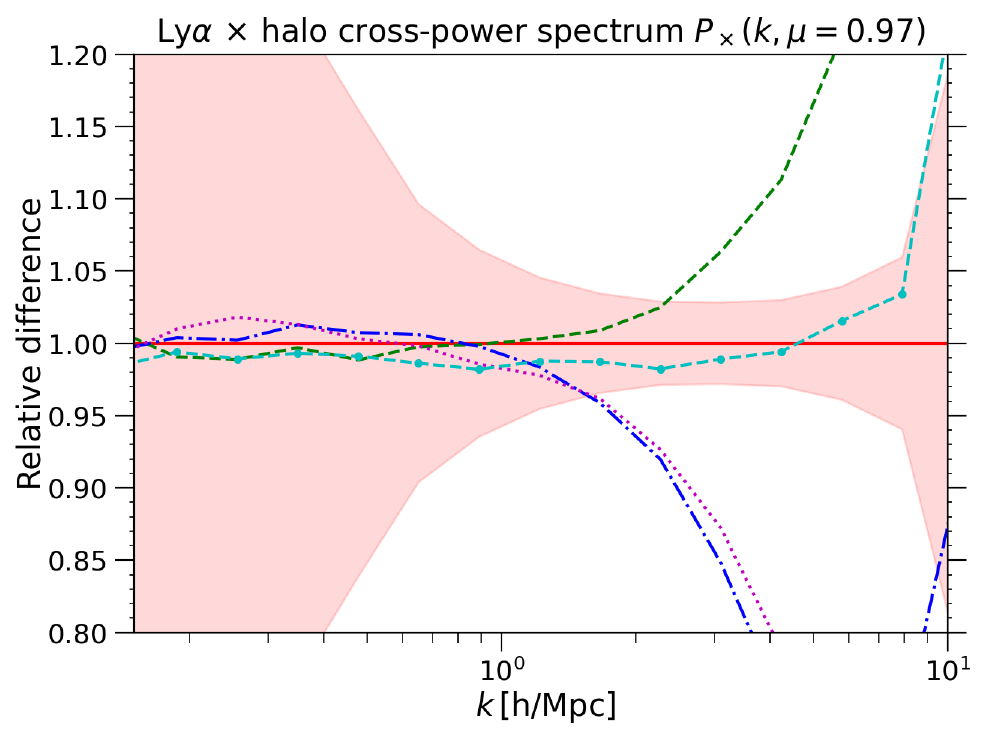}
 \end{minipage}
 \end{tabular}
 \caption{Comparison of the $\lya\,\times\,$halo cross-power spectrum among different astrophysical models, with shaded areas representing the $1\sigma$ error calculated according to equation (\ref{eq:error_cross}). 
 }
 \label{fig:comp_cross}
\end{figure*}

\subsubsection{Parameter estimation}
We fit the $\lya$ auto-power spectrum using equations (\ref{eq:autopower}) and (\ref{eq:arinyo}), where we treat seven parameters in equation (\ref{eq:params_auto}) as free parameters as described in Section~\ref{subsec:fitting_analysis}. Then we compare the best-fitting parameters among different astrophysical models. The results are summarized in Fig.~\ref{fig:compare}.
The linear bias parameter $b_F$ decides the amplitude of the power spectrum. We note that the fluctuation of the transmitted flux $\delta_F$ has anti-correlation with the density fluctuations of dark matter $\delta_\mathrm{DM}$ in the linear regime, resulting in a negative bias parameter. The RSD parameter $\beta_F$ characterizes $\mu$-dependence of the normalization of $\pf$. Mode estimation of these two parameters shows overall consistency within $\sim1\sigma$ error among different astrophysical models. We see some deviations of $b_F$ and $\beta_F$ in each model from the Fiducial model in Fig.~\ref{fig:compare}; however, these are not necessarily consistent with the deviations of power spectrum on large scales (at small $k$) in Fig.~\ref{fig:comp_auto}. This is because these linear parameters are degenerate with certain nuisance parameters in the fitting function (\ref{eq:arinyo}) as our analysis does not sufficiently include large-scale modes $>100\,\mpc$. For instance, $b_F$ is degenerate with $q_1$ and $\beta_F$ is degenerate with $a_v$. 
As a result, differences in the small-scale power spectrum can affect the linear parameters due to their degeneracy with nuisance parameters.

The parameter $q_1$ is enhanced for CW and Shield models over other models. This is consistent with the stronger power of these two models observed in the left panel of Fig.~\ref{fig:comp_auto}. The high $q_1$ value of Shield model can be explained as the clumping structure of HI gas resulting from the self-
shielding effects. 
The parameter $k_p$ is higher for the Shield model over other models, suggesting the Jeans smoothing occurs at slightly smaller scale than other models. However, the parameter $k_p$ just captures the $\mu$-independent suppression of the power. A reasonable explanation for the smaller suppression of power in the Shield model compared to the Fiducial model is that the HI gas remains un-ionized, leading to larger fluctuation amplitudes.
Parameters $\kvav$, $a_v$ and $b_v$ capture the $\mu$-dependence on small scales. We do not find any significant differences in $\kvav$ and $a_v$ among different models within $1\sigma$ errors, whereas the Shield model shows significantly lower $b_v$ value than other models. It implies a weaker $\mu$ dependence on small scales, although we do not delve into the physical origin of it, since those are marginalized and not that important for the latter cosmological parameter estimation. However, it is worth noting that if we aim to constrain the astrophysical model using $\lya$ power spectrum, $b_\nu$ parameter would be informative.

\begin{figure*}
\begin{center}
\includegraphics[width=0.9\linewidth]{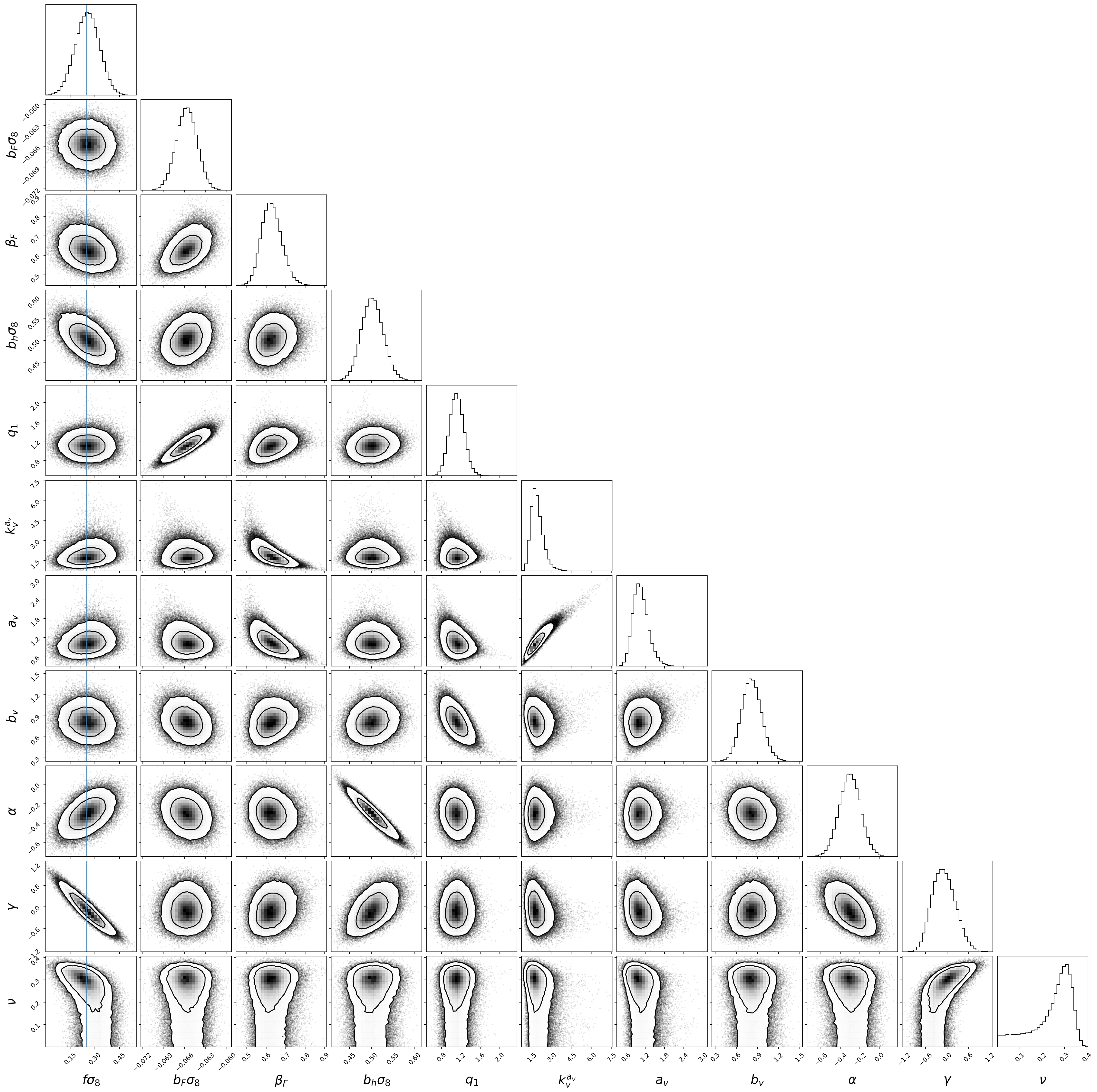}
 \caption{Posterior distribution for the $2\times 2$-pt fitting with {\it Givans} model. Contours show 68\% and 95\% confidence regions, and the solid line denotes the ground truth of $f\sigma_8$.
}
\label{fig:contour}
 \end{center}
\end{figure*}

%--------------------------------------------------------------------------------------
\subsection{Ly$\alpha\,\times$ halo cross-power spectrum}
%--------------------------------------------------------------------------------------

In Fig.~\ref{fig:crosspower}, we present the measured Ly$\alpha\,\times\,$halo cross-power spectra along with the theoretical best-fitting model. We show two different halo samples: one for  $M_h>10^{10.5}M_\odot$ (left) and the other is for $ M_h>10^{11.5}M_\odot$ (right). We note that $\pc$ is negative because of the anti-correlation between $\delta_F$ and halos.
The {\it Givans} model gives a very good fit to the measured power spectrum from our simulations up to $k=3\,\mpcinv$.

Figure~\ref{fig:comp_cross} shows a comparison of the cross power spectra among different astrophysical models relative to the Fiducial model. 
In the left panel, all models except for the Shield model are consistent with the Fiducial model on large scales, $k<1 \himpc$ within a few percent. On smaller scales, FG09 shows the least difference, $<5\%$ at $k<10\himpc$, while the CW model shows $10\%$ enhancement and the NoFB model shows $15\%$ suppression. 
In the right panel, all the models are consistent with the Fiducial model within the 1$\sigma$ error margin on large scales ($k<1\himpc$). On small scales, the CW and Shield models show suppression, while NoFB and FG09 models exhibit enhancement. All models, except FG09, display significant deviations greater than 5\%  at $k\gtrsim2\himpc$. In both panels, relative differences of $P_\times$ on small scales, $k>1\himpc$,  are larger compared to those of $P_F$ shown in Fig.~\ref{fig:comp_auto}.

\begin{figure*}
 \begin{tabular}{c|c}
  \begin{minipage}[t]{0.43\hsize}
   \centering
   \includegraphics[width=\columnwidth]{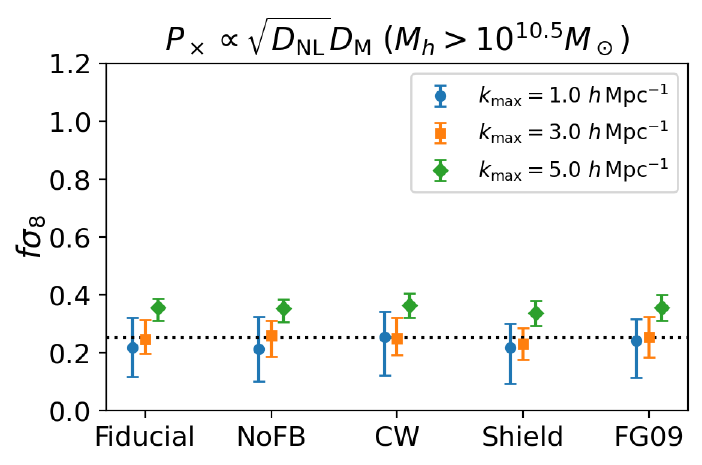}
   \subcaption{}
   \label{fig:constrain_dnldm_10.5}
  \end{minipage} &
  \begin{minipage}[t]{0.43\hsize}
   \centering
   \includegraphics[width=\columnwidth]{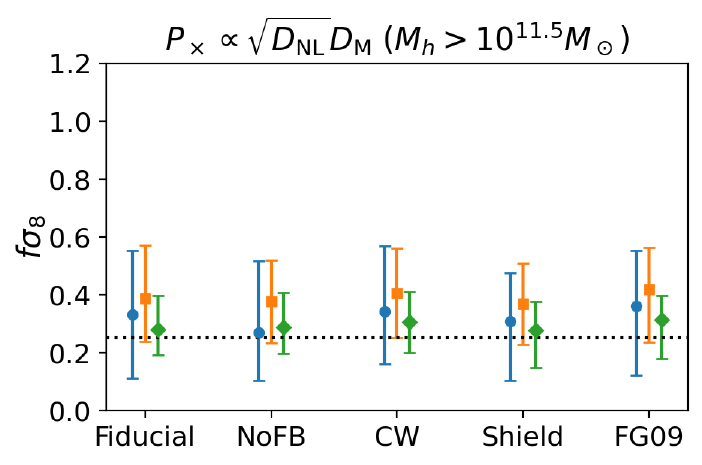}
   \subcaption{}
   \label{fig:constrain_dnldm_11.5}
  \end{minipage} \\
  
  \begin{minipage}[t]{0.43\hsize}
   \centering
   \includegraphics[width=\columnwidth]{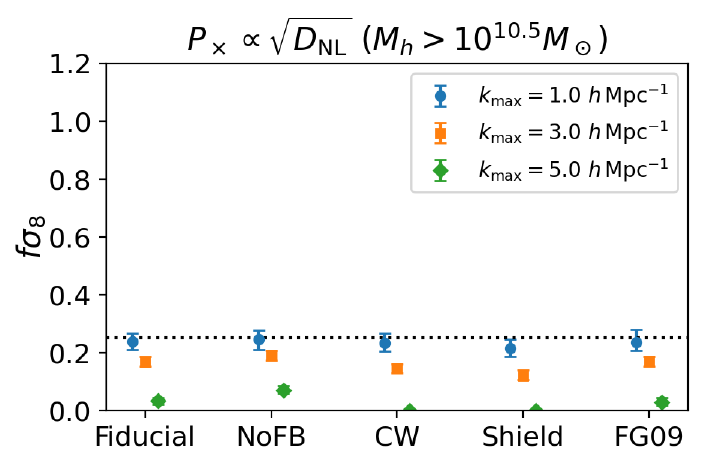}
   \subcaption{}
   \label{fig:constrain_dnl_10.5}
  \end{minipage} &
  \begin{minipage}[t]{0.43\hsize}
   \centering
   \includegraphics[width=\columnwidth]{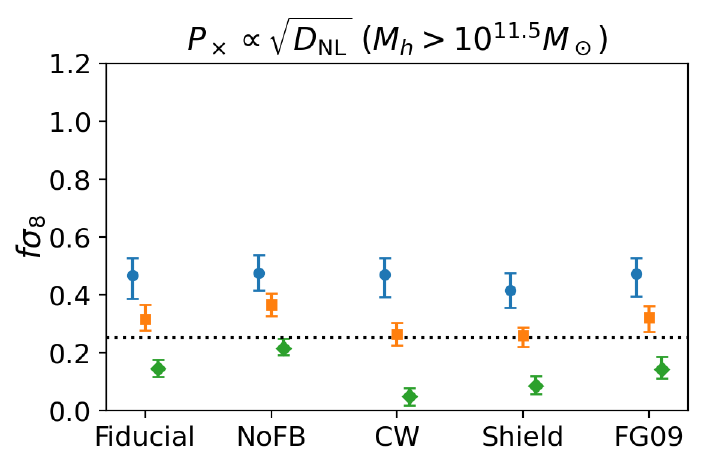}
   \subcaption{}
   \label{fig:constrain_dnl_11.5}
  \end{minipage} \\

  \begin{minipage}[t]{0.43\hsize}
   \centering
   \includegraphics[width=\columnwidth]{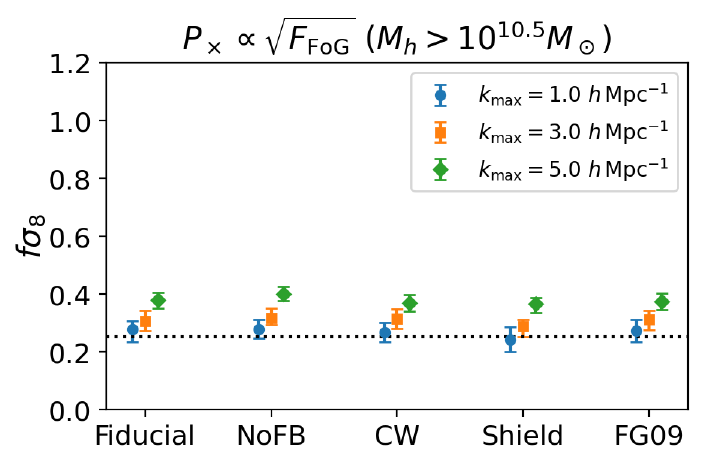}
   \subcaption{}
   \label{fig:constrain_fog_10.5}
  \end{minipage} &
  \begin{minipage}[t]{0.43\hsize}
   \centering
   \includegraphics[width=\columnwidth]{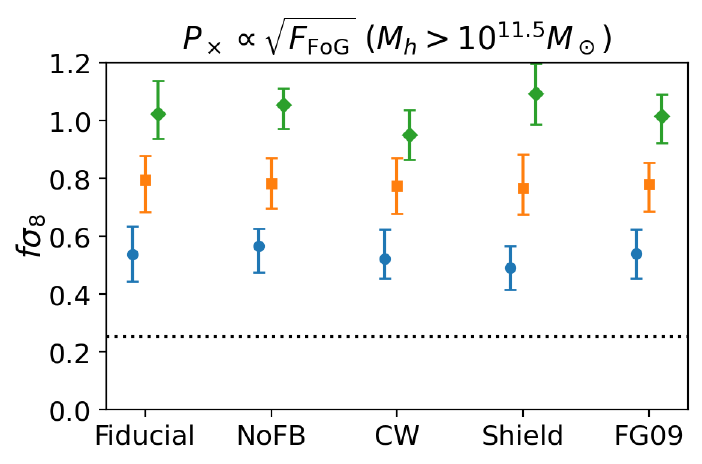}
   \subcaption{}
   \label{fig:constrain_fog_11.5}
  \end{minipage}

 \end{tabular}
 \caption{Constraints on $f\sigma_8$ from our five simulation models with varying maximum wave numbers $\kmax$. 
 The joint fitting of the $\lya$ auto-power spectrum and the cross-power spectrum with halos is performed, with 1$\sigma$ uncertainties derived through MCMC sampling, represented by the error bars. 
 The ground truth of $f\sigma_8$ at $z=3$ is indicated by the horizontal lines. In all panels, equation (\ref{eq:autopower}) is used for the $\pf$ model. The left and right panels correspond to different minimum halo masses $M_h$. 
 The differences between the $\pc$ models are shown from top to bottom:  panels ({\it a}) and ({\it b}) show results using equation (\ref{eq:crosspower}), panels ({\it c}) and ({\it d}) show results from equation (\ref{eq:pk_cross_dnl}), and panels ({\it e}) and ({\it f}) display the results from equation~(\ref{eq:Bourboux}).}
 \label{fig:constraints}
\end{figure*}

%--------------------------------------------------------------------------------------
\subsection{$f\sigma_8$ measurements with 2$\times$2-pt joint fitting}
\label{subsec:fs8constraints}
%--------------------------------------------------------------------------------------
In this section, we present our results for the joint constraints on the cosmological parameters and nuisance model parameters. Figure~\ref{fig:contour} shows the posterior distribution for the $2\times2$-pt fitting using {\it Givans} model. First, we discuss how $f\sigma_8$ estimation is influenced by the nuisance model parameters in the fitting function (\ref{eq:givans}). 
We observe weak correlations among model parameters, which makes the fitting stable, except for a few parameters. We also see that the parameter $\nu$ is unconstrained for smaller $\nu$.
We find a negative correlation between $b_h\sigma_8$ and $\alpha$, illustrating that $\alpha$ is independent of $\mu$ similarly to the linear bias factor. Additionally, we observe a negative correlation between $f\sigma_8$ and $\gamma$, which varies with $\mu^2$. These findings suggest that the precision of $f\sigma_8$ estimation could be improved by revising the model, which does not include these degenerate parameters.  Resolving the degeneracy between $b_h\sigma_8$ and $\alpha$, in addition to resolving the degeneracy between $f\sigma_8$ and $\gamma$, can improve the precision of the $f\sigma_8$ measurement because $b_h\sigma_8$ shows degeneracy with $f\sigma_8$,

Next, we explore the robustness of the $f\sigma_8$ measurement with respect to the choice of maximum wave numbers $\kmax$. Figure~\ref{fig:constraints} shows the best-fitting values and 68\% confidence regions for $f\sigma_8$ from simulations of different astrophysical models. The left (right) panel corresponds to the halo sample with $M_h>10^{10.5} (10^{11.5}) M_\odot$, respectively. In this section, we focus on the results from the {\it Givans} model (eq.~(\ref{eq:crosspower}) and (\ref{eq:givans})) shown in panels ({\it a}) \& ({\it b}). We observe that the error bars become smaller with higher $\kmax$ due to the increased number of $k$ modes available in both panels. 
In panel ({\it a}), the \textit{Givans} model reproduces the unbiased estimate of $f\sigma_8$ for $k_{\rm max} \leq 3 \himpc$, while for $k_{\rm max}=5 \himpc$, it shows highly inconsistent results with the ground truth, with more than 5 $\sigma$.
In panel ({\it b}), for $M_h<10^{11.5}M_\odot$, correct $f\sigma_8$ can be reproduced for all $k_{\rm max}$ with a slight overestimation at $k_{\rm max} = 3 \himpc$.
It is clearly seen that the error bars for $M_h>10^{11.5} M_\odot$ are larger than those for the $M_h>10^{10.5} M_\odot$ case, simply due to the smaller number of halos and larger shot noise contribution to the covariance. 
Comparing the results of different astrophysical models, we observe no significant biases in the $f\sigma_8$ estimation. This suggests that the {\it Givans} model determines $f\sigma_8$ values without being influenced by astrophysical uncertainty, despite more than 5\% differences at $k\gtrsim3\,\mpcinv$ as shown in Fig.~\ref{fig:comp_cross}.

We also performed the same analysis with the {\it Minimum} model 
%of Givans ($\pc \propto \dm$) model ({\it Givans} model without the $\sqrt{\dnl}$ term) 
and did not observe any significant differences compared to the {\it Givans} model.

\subsection{Comparing three fitting models; \mini model, {\it Givens} model and {\it FoG} model}
\label{subsec:compare_dnl_dnldm}
Finally, we compare the $f\sigma_8$ estimations across the fitting models. Figure~\ref{fig:constraints} shows the results from the {\it Givans} model (eq. \ref{eq:crosspower}; panels {\it a} \& {\it b}); \mini model (eq. \ref{eq:pk_cross_dnl}; panels {\it c} \& {\it d}), and {\it FoG} model (eq. \ref{eq:Bourboux}; panels {\it e} \& {\it f}), as summarised in Table~\ref{tab:fitting_models}. 

In panel ({\it c}), the $f\sigma_8$ estimations at $\kmax=1.0\,\mpcinv$ are consistent with the ground truth within $1\sigma$ error, except for the Shield model, while they become inconsistent at $\kmax=3.0\,\mpcinv$ and $\kmax=5.0\,\mpcinv$. In panel ({\it d}), using only massive halos ($\mh>10^{11.5}M_\odot$), $f\sigma_8$ estimations are consistent with ground truth only for two cases: CW and Shield model at $\kmax=3.0\,\mpcinv$. However, these two successes are not valid because they are inconsistent with ground truth even at $\kmax=1.0\,\mpcinv$. We also observe the significant difference across astrophysical models, which can be attributed to lack of degrees of freedom on small scales. The \mini model has no additional parameter in non-linear correction, therefore, it cannot express non-linearities of cross-power spectrum correctly. 

In panel ({\it e}), the $f\sigma_8$ estimations at $\kmax=1.0\,\mpcinv$ are consistent with the ground truth within $1\sigma$ error, while they become inconsistent at $\kmax=3.0\,\mpcinv$ and $\kmax=5.0\,\mpcinv$. In panel ({\it f}), the $f\sigma_8$ estimations are inconsistent with the ground truth in all cases. This is because the $\sqrt{\fog}$ term fails to capture the rapid suppression at $k=1-2\,\mpcinv$ in the $M_h >10^{11.5} M_\odot$ case (see the right panel of Figure~\ref{fig:crosspower}). The $\sqrt{\fog}$ term, a Lorentzian function, decreases more slowly than the Gaussian function at scales smaller than the pivot scale. As a result, the value of $\sigma_v$ must be large (causing earlier suppression) to fit the data points at $k>2\,\mpcinv$, which in turn makes $f\sigma_8$ larger to compensate for the early suppression. We also performed the same analysis using the $\pc \propto \sqrt{\dnl\fog}$ model and did not observe any significant differences compared to the {\it FoG} model.

As a result, the {\it Givans} model proves to be more  robust for estimating $f\sigma_8$ compared to the \mini and {\it FoG} models. However, the $1\sigma$ uncertainty with the {\it Givans} model is several times larger than that of the \mini and {\it FoG} models. For example, the {\it Givans} model yields $f\sigma_8=0.218^{+0.102}_{-0.102}$ for $M_h >10^{10.5} M_\odot$ at $\kmax=1.0\,\mpcinv$, while the \mini model provides $f\sigma_8=0.238^{+0.029}_{-0.029}$ and the {\it FoG} model gives $f\sigma_8=0.276^{+0.029}_{-0.043}$. 
This larger uncertainty arises because the {\it Givans} model accounts for correlations between $f\sigma_8$ and nuisance parameters, as discussed in Section~\ref{subsec:fs8constraints}. 

%======================================================================================
%
\section{Conclusions}\label{sec:conclusions}
%
%======================================================================================
In this study, we present measurements of the anisotropic $\lya$ power spectrum and its cross-power spectrum with halos derived from a suite of cosmological hydrodynamic simulations encompassing five different astrophysical models.
As shown in Fig.~\ref{fig:autopower}, the measured $\lya$ auto power spectrum is fairly well fitted by equation~(\ref{eq:autopower}) and the overall behaviour is consistent with the previous works \citep[e.g.][]{Arinyo2015, Givans2022}. 
The impact of astrophysical model differences on the power spectrum becomes noticeable at smaller scales. As depicted in Fig.~\ref{fig:comp_auto}, all models, except the Shield model, deviate from the Fiducial model by roughly $5\,\%$ up to $k=10\,\mpcinv$. We offer the physical interpretations of some of these behaviours, such as the enhanced power in the Shield model for $\mu=0.03$, which can be attributed to the clumping structure of HI gas due to the self-shielding effects.
In Fig.~\ref{fig:compare}, we compare the best-fitting parameters across the five models.   
For instance, the parameters $q_1$ and $k_p$ for the Shield model show significantly higher values compared to the other models, also due to the clumping structure of HI gas.

We also investigate the impact of different astrophysical models on the $\lya\,\times\,$halo cross-power spectrum, as shown in Fig.~\ref{fig:comp_cross}. We find that the relative differences of $P_\times$ among the different astrophysical models, exceeding 10\%, are larger compared to those of $P_F$, as shown in Fig.~\ref{fig:comp_auto}. However, as shown in Section~\ref{subsec:fs8constraints}, the estimation of $f\sigma_8$ using the {\it Givans} model is not impacted by these astrophysical uncertainties.

Next, we perform a joint fitting of the $\lya\,\times\,\lya$ auto- and the $\lya\,\times\,$halo cross-power spectrum using {\it Givans} model to estimate the growth rate parameter $f\sigma_8$. Fig.~\ref{fig:contour} shows the 2D contour plot of the parameter estimation resulting from the joint fitting. We observed the negative correlations between $b_h\sigma_8$ and $\alpha$, as well as between $f\sigma_8$ and $\gamma$, where $\alpha$ and $\gamma$ are the nuisance parameters in the {\it Givans} model. Refining this non-linear model of the cross-power spectrum to minimize the parameter degeneracies could improve the precision of the $f\sigma_8$ estimation. 

Finally, we conduct a fitting analysis under various conditions (varying $\kmax$ and $\mh$) to assess the accuracy of $f\sigma_8$ determination. As shown in Fig.~\ref{fig:constraints}, the \textit{Givans} model proves to be more robust to accurately reproduce $f\sigma_8$ in most cases, except when $M_h >10^{10.5} M_\odot$ at $\kmax=5.0\,\mpcinv$. In contrast, both the \mini and \textit{FoG} models do not provide accurate estimates. We also find that the \textit{Givans} and \textit{FoG} models can determine $f\sigma_8$ without being affected by astrophysical uncertainties. However, despite the overall success of the \textit{Givans} model, the statistical uncertainty in $f\sigma_8$ estimation is significantly larger than that of other models. Therefore, the \textit{FoG} model emerges as the best option for $M_h >10^{10.5} M_\odot$ sample at $\kmax=1.0\,\mpcinv$.

Our results support the assertion by \citet{Givans2022} that an additional correction term (eq.\ref{eq:givans}) is necessary when fitting $\lya$ cross-power spectrum with massive halos ($M_h >10^{11.5} M_\odot$). In contrast, \citet{Cuceu2021} adopted the {\it FoG} model in their forecast of RSD analysis and argued that using a smaller $r_\mathrm{min}$, the minimum separations used for fits, would improve $f\sigma_8$ constraints. However, since massive halos host quasars, including small-scale data could actually  hinder accurate estimates.
%======================================================================================
%
\section*{Acknowledgements}
%
%======================================================================================
%
%We thank the anonymous referee for detailed comments that helped to improve the manuscript significantly. 
We are grateful to Volker Springel for providing the original version of GADGET-3, on which the GADGET3-OSAKA code is based. Our numerical simulations and analyses were carried out on the XC50 systems at the Center for Computational Astrophysics (CfCA) of the National Astronomical Observatory of Japan (NAOJ), OCTOPUS and SQUID at the Cybermedia Center, Osaka University, and Oakforest-PACS at the University of Tokyo 
as part of the HPCI system Research Project (hp180063, hp190050, hp200041, hp230089, hp240141). This work is supported by the JSPS KAKENHI grant No. JP20H01932, JP21K03625, JP21H05454, JP23H00108 (A.N.), 19H05810, 20H00180, 24H00002, 24H00241 (K.N.), 21J20930, 22KJ2072 (Y.O.) and the JSPS International Leading Research (ILR) project, JP22K21349 (A.N. and K.N.). 
K.N. acknowledges the support from the Kavli IPMU, the World Premier Research Centre Initiative (WPI), UTIAS, the University of Tokyo.

%%%%%%%%%%%%%%%%%%%%%%%%%%%%%%%%%%%%%%%%%%%%%%%%%%



\bibliographystyle{mnras}
\bibliography{reference} % if your bibtex file is called example.bib

%\begin{thebibliography}{99}
%\end{thebibliography}

%%%%%%%%%%%%%%%%%%%%%%%%%%%%%%%%%%%%%%%%%%%%%%%%%%

%%%%%%%%%%%%%%%%% APPENDICES %%%%%%%%%%%%%%%%%%%%%

\appendix

\section{Some extra material}

\renewcommand{\arraystretch}{1.4}
\begin{table*}
 \caption{Bias parameters, non-linear fit parameters and chi-squared per degree of freedom values \csquared \,measured from  $\lya$ auto-power spectrum with $\kmax=10\,\mpcinv$ for five different astrophysical models.}
 \label{tab:params_auto}
 \centering
  \begin{tabular}{lccccccc}
  \hline
  \hline
  Parameters & Fiducial & NoFB & CW & Shield & FG09\\
  \hline
  $b_F$       & $-0.281^{+0.004}_{-0.004}$ & $-0.282^{+0.004}_{-0.003}$ & $-0.277^{+0.004}_{-0.004}$ & $-0.281^{+0.004}_{-0.004}$ & $-0.277^{+0.004}_{-0.004}$\\
  $\beta_F$     & $0.711^{+0.049}_{-0.049}$  & $0.688^{+0.049}_{-0.039}$  & $0.754^{+0.052}_{-0.052}$  & $0.727^{+0.043}_{-0.064}$  & $0.717^{+0.049}_{-0.049}$\\
  $q_1$       & $0.523^{+0.054}_{-0.065}$  & $0.489^{+0.064}_{-0.043}$  & $0.603^{+0.060}_{-0.060}$  & $0.602^{+0.069}_{-0.046}$  & $0.516^{+0.058}_{-0.058}$ \\
  $k_p$       & $8.708^{+0.173}_{-0.216}$  & $8.594^{+0.212}_{-0.170}$  & $8.698^{+0.202}_{-0.202}$ & $9.043^{+0.263}_{-0.263}$  & $8.789^{+0.233}_{-0.233}$ \\
  $k_v^{a_v}$ & $0.440^{+0.044}_{-0.035}$  & $0.455^{+0.032}_{-0.042}$  & $0.452^{+0.048}_{-0.032}$  & $0.454^{+0.050}_{-0.033}$  & $0.440^{+0.038}_{-0.038}$ \\
  $a_v$       & $0.222^{+0.044}_{-0.044}$  & $0.234^{+0.046}_{-0.046}$  & $0.194^{+0.047}_{-0.047}$  & $0.201^{+0.042}_{-0.042}$  & $0.217^{+0.036}_{-0.046}$ \\
  $b_v$       & $1.447^{+0.058}_{-0.046}$  & $1.504^{+0.061}_{-0.049}$  & $1.405^{+0.039}_{-0.049}$  & $1.201^{+0.046}_{-0.057}$  & $1.453^{+0.045}_{-0.056}$ \\
  \hline
  \csquared\ (best fit)   & $1.033$ & $1.049$ & $0.984$ & $1.106$ & $1.044$ \\
  
 \hline
\end{tabular}
\end{table*}

\begin{table*}
 \caption{Bias parameters, non-linear fit parameters and chi-squared per degree of freedom values measured by joint analysis of $\lya\,\times\,\lya$ and $\lya\,\times\,$halo cross-power spectrum with $\mh=10^{10.5}M_\odot$ using {\it Givans} fitting model. Here, we show results from Fiducial model.}
 \label{tab:params_k10.5}
 \centering
  \begin{tabular}{lccc}
  \hline
  \hline
  Parameters & $\kmax=1\,\mpcinv$ & $\kmax=3\,\mpcinv$ & $\kmax=5\,\mpcinv$ \\
  \hline
  $b_F\sigma_8$  & $-0.068^{+0.002}_{-0.002}$ & $-0.066^{+0.001}_{-0.001}$ & $-0.069^{+0.001}_{-0.001}$ \\
  $\beta_F$      & $0.570^{+0.031}_{-0.046}$  & $0.623^{+0.050}_{-0.050}$  & $0.584^{+0.031}_{-0.031}$  \\
  $b_h\sigma_8$  & $0.443^{+0.051}_{-0.038}$  & $0.497^{+0.027}_{-0.020}$  & $0.514^{+0.015}_{-0.015}$ \\
  $f\sigma_8$    & $0.218^{+0.102}_{-0.102}$  & $0.246^{+0.068}_{-0.051}$  & $0.355^{+0.031}_{-0.046}$ \\
  $q_1$          & $0.736^{+0.233}_{-0.233}$  & $1.097^{+0.139}_{-0.139}$  & $0.751^{+0.068}_{-0.068}$  \\
  $k_v^{a_v}$    & $1.178^{+4.334}_{-0.619}$  & $1.772^{+0.397}_{-0.397}$  & $1.447^{+0.192}_{-0.128}$  \\
  $a_v$          & $2.031^{+4.687}_{-0.937}$  & $0.966^{+0.212}_{-0.141}$  & $0.915^{+0.067}_{-0.067}$  \\
  $b_v$          & $1.406^{+4.062}_{-1.562}$  & $0.796^{+0.131}_{-0.087}$  & $0.794^{+0.062}_{-0.062}$  \\
  $\alpha$       & $-0.033^{+0.415}_{-0.311}$ & $-0.297^{+0.088}_{-0.088}$ & $-0.294^{+0.051}_{-0.068}$ \\
  $\gamma$       & $-0.068^{+0.981}_{-0.588}$ & $-0.152^{+0.314}_{-0.235}$ & $-0.714^{+0.143}_{-0.096}$  \\
  $\nu$          & $0.497^{+0.137}_{-0.069}$  & $0.304^{+0.052}_{-0.078}$  & $0.224^{+0.009}_{-0.014}$  \\
  \hline
  \csquared\ (best fit)      & $0.435$ & $0.458$ & $1.318$ \\
  
 \hline
\end{tabular}
\end{table*}

\begin{table*}
 \caption{Bias parameters, non-linear fit parameters and chi-squared per degree of freedom values measured by joint analysis of $\lya\,\times\,\lya$ and $\lya\,\times\,$halo cross-power spectrum with $\mh=10^{11.5}M_\odot$ using {\it Givans} fitting model. Here, we show only results from Fiducial model.}
 \label{tab:params_k11.5}
 \centering
  \begin{tabular}{lccc}
  \hline
  \hline
  Parameters & $\kmax=1\,\mpcinv$ & $\kmax=3\,\mpcinv$ & $\kmax=5\,\mpcinv$ \\
  \hline
  $b_F\sigma_8$  & $-0.068^{+0.002}_{-0.003}$ & $-0.067^{+0.001}_{-0.001}$ & $-0.072^{+0.001}_{-0.001}$ \\
  $\beta_F$      & $0.562^{+0.032}_{-0.032}$  & $0.607^{+0.047}_{-0.047}$  & $0.546^{+0.021}_{-0.032}$  \\
  $b_h\sigma_8$  & $0.682^{+0.078}_{-0.078}$  & $0.791^{+0.068}_{-0.051}$  & $0.731^{+0.045}_{-0.045}$ \\
  $f\sigma_8$    & $0.331^{+0.221}_{-0.221}$  & $0.378^{+0.180}_{-0.144}$  & $0.278^{+0.117}_{-0.088}$ \\
  $q_1$          & $0.729^{+0.194}_{-0.194}$  & $0.970^{+0.109}_{-0.109}$  & $0.474^{+0.059}_{-0.059}$  \\
  $k_v^{a_v}$    & $1.200^{+5.249}_{-0.618}$  & $1.744^{+0.260}_{-0.520}$  & $1.188^{+0.210}_{-0.210}$  \\
  $a_v$          & $2.031^{+4.375}_{-1.250}$  & $1.015^{+0.198}_{-0.198}$  & $1.006^{+0.102}_{-0.102}$  \\
  $b_v$          & $1.406^{+3.750}_{-1.562}$  & $0.924^{+0.117}_{-0.117}$  & $1.015^{+0.074}_{-0.074}$  \\
  $\alpha$       & $-0.767^{+0.617}_{-0.462}$ & $-1.121^{+0.168}_{-0.168}$ & $-0.893^{+0.213}_{-0.160}$  \\
  $\gamma$       & $0.623^{+1.219}_{-0.732}$  & $0.569^{+0.395}_{-0.527}$  & $1.030^{+0.464}_{-0.348}$ \\
  $\nu$          & $0.092^{+0.517}_{-0.111}$  & $0.400^{+0.034}_{-0.068}$  & $0.522^{+0.031}_{-0.031}$ \\
  \hline
  \csquared\ (best fit)     & $0.401$ & $0.568$ & $1.575$ \\
  
 \hline
\end{tabular}
\end{table*}

%%%%%%%%%%%%%%%%%%%%%%%%%%%%%%%%%%%%%%%%%%%%%%%%%%

% Don't change these lines
\bsp	% typesetting comment
\label{lastpage}
\end{document}